\documentclass[sigconf, nonacm]{acmart}

\newcommand\vldbdoi{XX.XX/XXX.XX}
\newcommand\vldbpages{XXX-XXX}
\newcommand\vldbvolume{14}
\newcommand\vldbissue{1}
\newcommand\vldbyear{2020}
\newcommand\vldbauthors{\authors}
\newcommand\vldbtitle{\shorttitle} 
\newcommand\vldbavailabilityurl{URL_TO_YOUR_ARTIFACTS}
\newcommand\vldbpagestyle{plain} 

\def\showcomments{1}
\def\removespace{1}

\def\camerareadyversion{0}

\usepackage{tikz}
\usepackage{amsmath}

\usepackage{pifont}% http://ctan.org/pkg/pifont

\usepackage{booktabs}
\usepackage{filecontents}

\usepackage{paralist}
\usepackage{xspace}
\usepackage{tabularx}
\usepackage{subfigure}
\usepackage{multirow}
\usepackage{enumitem}
\usepackage[utf8]{inputenc}

\usepackage{cleveref}

\crefformat{section}{\S#2#1#3}
\crefformat{subsection}{\S#2#1#3}
\crefformat{subsubsection}{\S#2#1#3}

\renewcommand\footnotetextcopyrightpermission[1]{}
\date{}

\newcommand{\mycaption}[2]{\caption{\textbf{#1}. {#2}}}
\newcommand{\sref}[1]{\S\ref{#1}}
\newcommand{\vheading}[1]{\vspace{0.05in}\noindent\textbf{#1}}
\newcommand{\viheading}[1]{\vspace{0.05in}\noindent\emph{#1}}

\newcommand{\eg}{\textit{e.g.,}\xspace}

\newcommand{\myx}{$\times$\xspace}
\newcommand{\vtt}[1]{\texttt{#1}\xspace}

\newcommand{\mmap}{memory-map\xspace}

\newcommand{\mmaps}{{memory-maps}\xspace}

\newcommand{\vio}{{I/O}\xspace}

\if\showcomments 1
\newcommand{\souji}[1]{{\textcolor{orange}{Soujanya: #1}}}
\newcommand{\skl}[1]{{\textcolor{purple}{Sekwon: #1}}}
\newcommand{\rohan}[1]{{\textcolor{blue}{Rohan: #1}}}
\newcommand{\vijay}[1]{{\textcolor{red}{Vijay: #1}}}
\else
\newcommand{\souji}[1]{}
\newcommand{\skl}[1]{}
\newcommand{\rohan}[1]{}
\newcommand{\vijay}[1]{}
\fi

\newcommand{\sysname}{{\textsc{SKYE}}\xspace}

\newcommand{\nvdimms}{{NVDIMMs}\xspace}
\newcommand{\nvdimm}{{NVDIMM}\xspace}
\newcommand{\worker}{{{worker}}\xspace}
\newcommand{\workers}{{{workers}}\xspace}

\newcommand{\nvlog}{\textsc{nvLOG}\xspace}
\newcommand{\nvlogs}{\textsc{nvLOGs}\xspace}

\usepackage{ifthen}
\newboolean{publicversion}
\setboolean{publicversion}{false}
\ifthenelse{\boolean{publicversion}}{
  \newcommand{\grumbler}[3]{}
}{
  \newcommand{\grumbler}[3]{\textcolor{#3}{\bf #1: #2}}
}

\begin{document}

% \title{\Large{SKYE: Scalably Saturating Persistent-Memory
%   Write Bandwidth with Carefully-Controlled I/O}}
% \title{\Large{SKYE: Scalable, High-Performance Persistent Memory Key-Value Store}}
% \title{\Large{SKYE: Persistent Memory Key-Value Store with Fine-Grained Control over I/O}}
% \title{\Large{Designing Key-Value Stores with Fine-Grained Control over Persistent Memory Accesses}}
% \title{\Large{High-Performance, Scalable Key-Value Store with Software-Managed Control to Persistent Memory}}

% \title{\Large{High-Performance Key-Value Store with Software-Managed Control over Persistent Memory}}
% \title{\sysname: Improving Key-Value Store Performance with Fine-Grained Control over Persistent Memory}

% \title{\Large{\sysname: A Systematic Approach to Scalably Saturating Persistent Memory Write Bandwidth}}

\title{\Large{\sysname: Write-Optimized Key-Value Store with Fine-Grained Control over\\ Persistent Memory Accesses}}

%%
%% The "author" command and its associated commands are used to define the authors and their affiliations.
\author{Soujanya Ponnapalli}
\affiliation{%
  \institution{\small{University of Texas at Austin}}
}
\email{soujanyap95@gmail.com}
% \if\camerareadyversion 1
% \email{soujanyap95@gmail.com}
% \fi

\author{Sekwon Lee}
\affiliation{%
  \institution{\small{University of Texas at Austin}}
}
\email{sekwonlee90@gmail.com}
% \if\camerareadyversion 1
% \email{sekwonlee90@gmail.com}
% \fi

\author{Rohan Kadekodi}
\affiliation{%
  \institution{\small{University of Texas at Austin}}
}
\email{kadekodirohan@gmail.com}
% \if\camerareadyversion 1
% \email{kadekodirohan@gmail.com}
% \fi

\author{Vijay Chidambaram}
\affiliation{%
  \institution{\small{University of Texas at Austin}}
}
\email{vijayc@utexas.edu}
%\affiliation{%
%  \institution{VMware Research Group}
%}
% \if\camerareadyversion 1
% \fi

\author{}
\affiliation{}

\author{}
\affiliation{}

\begin{abstract}
State-of-the-art key-value stores built for persistent memory (PM) provide low latency as they allow application threads to directly access the data on PM and rely on hardware to manage multiple non-volatile DIMMs (\nvdimms). 
While this provides low latency, it results in low throughput and scalability. Performance degrades because PM hardware requires fine-grained control over PM accesses; for example, throughput degrades if too many threads write to PM concurrently.  
%Since PM key-value stores no longer control how many concurrent threads access an \nvdimm and the data placement on individual \nvdimms, they obtain a small fraction of PM bandwidth and trade off high throughput and scalability.

We present \sysname, a write-optimized PM key-value store that achieves high throughput and scalability.
\sysname builds on the central idea of maintaining \emph{fine-grained control over all PM accesses} and obtains high PM write-bandwidth utilization.
To achieve this, \sysname deviates from current practice and provides \emph{indirect access} to applications; 
applications send requests to \sysname, which uses dedicated threads to access PM on their behalf. 
%\sysname provides indirect-access to applications.
%First, \sysname uses dedicated \emph{\worker} threads and
 % decouples PM accesses from application threads.
%Second, 
Instead of relying on hardware-managed PM,
  \sysname controls how data is placed on individual \nvdimms.
\sysname leverages multiple media to avoid overloading PM and
  limits remote NUMA accesses for scalable throughput.
We show that on a single \nvdimm, 
  \sysname outperforms state-of-the-art PM stores
  by 2.5--5\myx on the standard Yahoo Cloud Serving Benchmark (YCSB).
With four \nvdimms across four NUMA nodes,
  \sysname obtains $\approx$86\% of PM write bandwidth, and its
  write throughput scales by 3.9\myx.
\if \removespace 1
\vspace{-20pt}
\fi
\end{abstract}
% and can scale to multiple \nvdimms across non-uniform memory access (NUMA) nodes.
% \sysname trades off low latency for high throughput and presents a unique point in the design space of PM stores.

\settopmatter{authorsperrow=4}
\settopmatter{printfolios=true}
\maketitle
\pagestyle{plain}

\if\camerareadyversion 0
\setcounter{page}{1}
\fi

\pagestyle{\vldbpagestyle}
\if\camerareadyversion 1
\begingroup\small\noindent\raggedright\textbf{PVLDB Reference Format:}\\
\vldbauthors. \vldbtitle. PVLDB, \vldbvolume(\vldbissue): \vldbpages, \vldbyear.\\
\href{https://doi.org/\vldbdoi}{doi:\vldbdoi}
\endgroup
\begingroup
\renewcommand\thefootnote{}\footnote{\noindent
This work is licensed under the Creative Commons BY-NC-ND 4.0 International License. Visit \url{https://creativecommons.org/licenses/by-nc-nd/4.0/} to view a copy of this license. For any use beyond those covered by this license, obtain permission by emailing \href{mailto:info@vldb.org}{info@vldb.org}. Copyright is held by the owner/author(s). Publication rights licensed to the VLDB Endowment. \\
\raggedright Proceedings of the VLDB Endowment, Vol. \vldbvolume, No. \vldbissue\ %
ISSN 2150-8097. \\
\href{https://doi.org/\vldbdoi}{doi:\vldbdoi} \\
}\addtocounter{footnote}{-1}\endgroup
\fi

\if\camerareadyversion 1
\ifdefempty{\vldbavailabilityurl}{}{
\vspace{.3cm}
\begingroup\small\noindent\raggedright\textbf{PVLDB Artifact Availability:}\\
The source code, data, and/or other artifacts have been made available at \url{\vldbavailabilityurl}.
\endgroup
}
\fi

\section{Introduction}
\label{sec:Introduction}

Persistent Memory (PM) refers to storage-class memory
  that offers durability and byte-addressability.
Intel's Optane DC Persistent Memory Module~\cite{intel_dcpmm}
  was the first commercially-available media of this kind;
  other companies are also working on persistent
  memory technologies (\eg PCM~\cite{burr2010phase},
  STT-MRAM~\cite{apalkov2013spin}, Memristor~\cite{yang2013memristive},
  Samsung's memory-semantic SSD~\cite{samsung_cxl_ssd}).
PM has low latency (similar to DRAM) and high bandwidth (10\myx compared to modern SSDs)~\cite{yang2020empirical}.
The PM capacity and bandwidth on a single machine
  will increase with the emerging Compute Express Link (CXL)~\cite{cxl},
  a cache-coherent interconnect for processors~\cite{sapphire_rapids} and memory devices;
  PM will also be accessible over the PCIe bus as a Type3 device~\cite{cxl, cxl2.0,jung2022hello}.
% The high bandwidth and low latency of PM
%   make it a promising building block for large-scale persistent
%   key-value stores~\cite{rocksdb, ghemawat2011leveldb}.
The large capacity, low latency, and high bandwidth of PM
  make it a promising building block for large-scale persistent
  key-value stores.

\begin{figure}
\centering
\small
\includegraphics[width=\columnwidth]{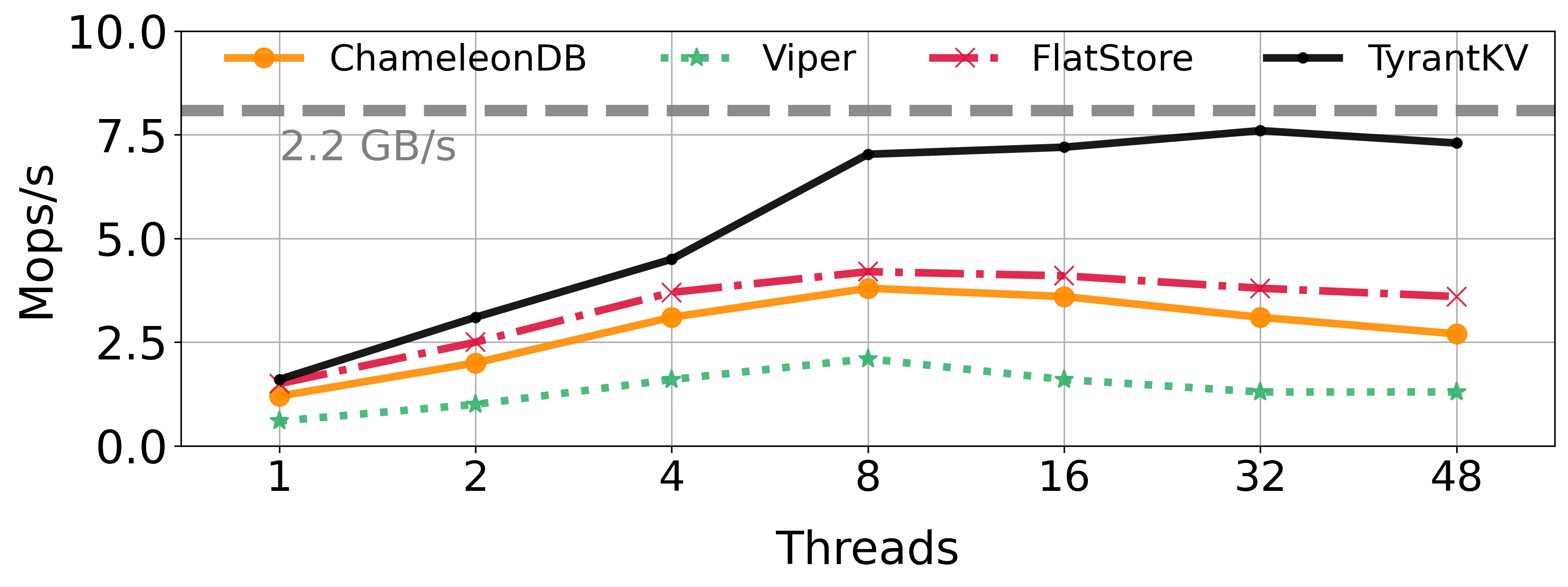}
\if\removespace 1
\vspace{-20pt}
\fi
\mycaption{Throughput and scalability}{Existing PM stores have low write throughput which drops beyond 8 threads. \sysname achieves high and scalable write throughput.}
\label{fig:scalability-others}
\if\removespace 1
\vspace{-15pt}
\fi
\end{figure}

% lets motivate key-value stores and write optimization
Many widely-used \vio-intensive applications such as Apache Flink~\cite{flink}, DGraph~\cite{flink}, CockroachDB~\cite{cockroachdb}, and Pinterest~\cite{pinterest-rocksdb} are built on top of key-value stores. 
These applications use in-memory caches like Memcached~\cite{memcached} to get high read throughput,
and require the key-value store to provide durable writes at high throughput.
As a result, the most widely-used key-value stores such as LevelDB~\cite{ghemawat2011leveldb} and RocksDB~\cite{rocksdb} are write-optimized.
These key-values stores are able to write data at close to the write bandwidth provided by magnetic hard drives and solid state drives. 

\if 0
Building a scalable, high throughput PM key-value store has
  the potential to impact a large number of widely-used
  \vio-intensive applications and platforms such as Apache Flink, DGraph, and Pinterest~\cite{cockroachdb, dgraph, uber-rocksdb, flink, pinterest-rocksdb, iron-rocksdb,samza}.
Today, high throughput write-optimized key-value stores~\cite{rocksdb, ghemawat2011leveldb} are key components in distributed systems~\cite{chang2008bigtable,annamalai2015zippydb,sivasubramanian2012amazon,burrows2006chubby,dean2008mapreduce}
  and cloud platforms~\cite{amazon_s3,google_gcp}.
These applications rely on the underlying
  key-value stores to obtain scalable and high throughput. 
% Lets cite more here!
\fi

This is in sharp contrast to state-of-the-art PM key-value stores that achieve low latency and good read performance, but offer poor write throughput~\cite{zhang2021chameleondb,wang2022pacman,benson2021viper,chen2020flatstore}. 
Figure-\ref{fig:scalability-others} shows the write throughput of several state-of-the-art PM key-value stores on a write only workload. 
The best-performing key-value store, FlatStore~\cite{chen2020flatstore}, only achieves about 45\% of the write bandwidth of a single PM \nvdimm;
furthermore, write throughput does not scale beyond eight threads. 
%%%
\if 0
Unfortunately, state-of-the-art PM key-value stores~\cite{zhang2021chameleondb,wang2022pacman,benson2021viper,chen2020flatstore}
  suffer from poor performance.
We observe that their performance degrades beyond 8 application threads (Figure-\ref{fig:scalability-others}).
Furthermore, these PM stores utilize $<$45\% of PM write
  bandwidth and $<$10\% of PM read bandwidth (\sref{sec:Background})
  with varying access sizes (256B--1kB).
\fi
%%
The low throughput and scalability of existing PM stores 
  is a fundamental consequence of their latency-oriented designs.
%We discuss why obtaining low latency for applications and high bandwidth utilization are competing goals with PM.
% This tradeoff between low latency and high throughput is due to two ?????.
% discuss is a bit much here because the discussion is not very deep

% getting low latency
To achieve low latency, current PM key-value stores allow applications to directly read and write to PM. 
They minimize additional accesses in the critical path of processing requests. 
They allow the hardware to manage multiple \nvdimms and present a unified device. 
However, through this approach, current PM stores \emph{give up control}: 
applications decide how many reads and writes happen concurrently to an \nvdimm,
and the hardware controls the striping of data across different \nvdimms. 

% getting high throughput
Unfortunately, this approach is directly at odds with getting high throughput, especially write throughput. 
Recent PM studies~\cite{izraelevitz2019basic,yang2020empirical,daase2021maximizing}
    and our analysis (\sref{sec:Analysis}) highlight that PM stores must
(R1) limit the number of concurrent reads and writes to \nvdimms and
(R2) manage data placement across individual \nvdimms,
instead of relying on memory-controllers,
to obtain high throughput.
If PM stores do not carefully control PM accesses, then the memory-controller gets overloaded, which reduces PM bandwidth and lowers the peak throughput of PM stores.

% focus on bw
While the focus of current PM key-value stores has been on latency, PM offers an order of magnitude higher bandwidth than existing persistent storage. 
It is important to maximize the utilization of PM write bandwidth so that applications such as Flink~\cite{flink} and CockroachDB~\cite{cockroachdb} can achieve durability at high throughput. 
In this work, we explore this part of the design space, building a write-optimized PM key-value store.

\vheading{Our contribution}.
We present \sysname, a novel write-optimized key-value store for PM.
The primary goal of \sysname is to provide high write throughput in a scalable fashion across multiple \nvdimms and multiple NUMA nodes. 
\sysname strikes a balance between write throughput and read throughput; for example, though read performance is not the primary goal, 
\sysname achieves read performance comparable or better than current PM key-value stores.
\sysname is targeted at applications that use key-value stores such as RocksDB for persisting large amounts of data;
\sysname is not a good fit for applications that require low latency read and write operations. 

%%%%%%%
\if 0
\vheading{Our contribution}.
We present \sysname,
  a novel PM key-value store that maintains
  fine-grained control over all PM accesses and provides
  high throughput to applications instead of low latency.
\sysname takes advantage of the high bandwidth of PM.
The high throughput of \sysname remains stable with increasing
  number of application threads (Figure-\ref{fig:scalability-others}).
On a single \nvdimm, \sysname outperforms state-of-the-art PM stores on all YCSB workloads by 2--5\myx.
\sysname achieves scalable performance to \nvdimms across
 non-uniform memory access (NUMA) nodes.
 \sysname is designed to write sequentially to \nvdimms and is write-optimized.
We observe that random accesses impact write throughput by $\approx$5\myx while
reducing read throughput by $<$20\%. Thus, we design a write-optimized PM store.
\rohan{Why don't we say that \sysname is a write optimized key-value store that is aimed at saturating PM write bandwidth, without compromising on the read performance. And in the results, we can say that we outperform others by 2-5\myx on write-heavy workloads and 1-2\myx on read-heavy workloads?. We can then say that we observe that building a write optimized key-value store helps get high write bandwidth as well as good read throughput. On the other hand, building a read-optimized key-value stores severely compromises write throughput.}
\fi
%%%%%

%vc: flow from here on is not great

Central to the design of \sysname is the concept of \emph{controlling accesses to PM}.
This concept is manifested in a number of ways.
First, contrary to current PM key-value stores, \sysname does not allow direct access to PM.
Application requests are placed in a queue, and \sysname worker threads dequeue and handle requests, performing \vio directly on PM.
This is important because unfettered access to PM can reduce PM bandwidth utilization and application throughput;
\sysname carefully controls how many threads are accessing \nvdimm at any given time.
This way, the \nvdimm controller is never overloaded with requests. 

Second, \sysname individually manages each PM \nvdimm, instead of using hardware provided \emph{interleaving}. 
This again goes back to the notion of control, of carefully placing data and controlling read and write access to each \nvdimm. 
We observe that managing individual \nvdimms provides $\approx$20\%
  higher throughput for reads and writes.
With hardware interleaving, every 4KB block would be spread across the available \nvdimms; in contrast, \sysname distributes data at the granularity of each write request.
%This is important since key-value pairs tend to be small for many applications, so the request-level distribution of \sysname provides better performance than distributing every 4KB of key-value pairs. 
\sysname associates each worker thread with a data and metadata log on \nvdimm,
  and distributes write requests uniformly across available workers.
\sysname ensures that the workers are tied to \nvdimms in the same NUMA node, to minimize expensive cross-NUMA-node traffic.

Third, \sysname uses PM to store data and metadata (in the form of logs), but uses DRAM to store indexes for looking up keys. 
This design prevents interference between the index lookup traffic and the writes on PM, maintaining high write throughput. 
\sysname periodically checkpoints the index to disk drives, and recovers the index if there is a crash or a power failure. 
The use of multiple storage media is an important factor in \sysname achieving good write performance without sacrificing read performance.

\vheading{Trade offs and limitations}.
The architecture of \sysname has two limitations. 
First, \sysname fundamentally trades off low latency for high throughput.
The indirect-access architecture allows high utilization of PM write bandwidth ($\approx$ 88\%), but also increases latency to several microseconds. 
Thus, \sysname is a good match for applications seeking high write throughput, but not low latency. 
Second, \sysname incurs high CPU utilization, since \sysname worker threads access PM on behalf of applications. 
To achieve high bandwidth utilization, a significant number of worker threads are required. 
To lower its CPU utilization, \sysname allows concurrent reads and writes;
this is in contrast to prior work that recommends only reads or writes at a given time~\cite{daase2021maximizing,yang2020empirical}. 
Our analysis finds that mixed \vio to \nvdimms achieves similar throughput with fewer threads.
Each \worker in \sysname processes reads and writes requests, and processes write requests in batches, to achieve high throughput and lower its CPU utilization.

\vheading{Evaluation}.
We evaluate \sysname using the YCSB benchmark suite~\cite{cooper2010benchmarking}.
We compare \sysname against the state-of-the-art PM key-value stores
  like Viper~\cite{benson2021viper}, ChameleonDB~\cite{zhang2021chameleondb},
  and FlatStore~\cite{chen2020flatstore}.
On a single \nvdimm,
  across all YCSB workloads,
  \sysname outperforms FlatStore by 2\myx,
  Viper by 3--5\myx,
  and ChameleonDB by 3\myx.
\sysname achieves 88\% of PM write bandwidth
  and it scales to \nvdimms across NUMA nodes.
We show that under low request load and by disabling request batching, \sysname achieves comparable latencies as existing
  PM stores (while trading off high PM bandwidth utilization).
On single NUMA node,
  \sysname outperforms these PM stores by $\approx$2\myx
  on write-dominant workloads and achieves
  comparable throughput on read-dominant workloads (\sref{sec:eval}).

This work shows some of the fundamental trade-offs that need to be made in designing a write-optimized key-value store for PM.
The design is quite different from current PM key-value stores, and the design choices should prove instructive to practitioners building high-throughput storage systems for PM or similar byte-addressable media.
We observe that though \sysname is write-optimized, it achieves comparable or better read throughput than current PM stores;
we believe building a read-optimized key-value store that achieves good write throughput is significantly harder. %rohan's nice observation

This paper makes the following contributions:
\begin{itemize}[leftmargin=*,itemsep=0pt]
  \item An analysis of existing PM stores (\sref{sec:Background})
  and an empirical study on PM (\sref{sec:Analysis})
  which highlights
  the need for fine-grained control over PM
  to achieve high throughput
  % \item An empirical analysis of PM performance under different
  %   configurations, resulting in a set of principles for
  %   achieving high PM write-bandwidth utilization 
  \item The design (\sref{sec:design}) and implementation
    (\sref{sec:Implementation}) of \sysname, a novel PM key-value
    store that obtains high PM bandwidth utilization
    and scalable throughput
  \item An empirical evaluation demonstrating that our
  approach achieves good throughput on a range of workloads (\sref{sec:eval})
\end{itemize}

\section{Background and Motivation}
\label{sec:Background}

We present an overview of persistent memory (PM) and describe
its unique characteristics.
We discuss the performance and
  scalability of existing PM stores and outline
  their design choices.

\subsection{Persistent Memory}

PM refers to a class of memory technologies that provide durable, byte-addressable media.
PM is available as individual non-volatile DIMMs (\nvdimms)
  which can be directly connected to the memory-bus like DRAM.
PM is cheaper and denser than DRAM;
  a single server with 4 non-uniform memory access (NUMA) nodes
  can support up to 12 TB of PM (six 512 GB-sized \nvdimms per node).
PM has desirable qualities like high capacity and
  low cost-per-GB relative to DRAM, and has lower latency and higher bandwidth compared to solid state drives and magnetic hard drives.
Thus, PM is a promising building block for key-value stores.

% asymmetric read/write bw
\vheading{Latency and bandwidth}.
PM has asymmetric latency and bandwidth for reads and writes.
Loads on PM incur 2--3\myx higher latency
  than on DRAM, while stores incur similar latency on PM
  and DRAM~\cite{intel_dcpmm}.
The write bandwidth of a single \nvdimm
  is 2.3 GB/s ($1/6^{th}$ of DRAM's write bandwidth)  and its read bandwidth
  is 6.6 GB/s ($1/3^{rd}$ of DRAM's read bandwidth)~\cite{izraelevitz2019basic}.
PM has high \vio bandwidth relative to disks (10\myx compared to modern SSDs).
However PM bandwidth is sensitive to
 the access patterns, sizes, and the number of
 concurrent operations. 
% Thus, PM stores
%   must effectively utilize the low \vio bandwidth of PM
%   to achieve high performance.

\vheading{AppDirect mode}.
PM can be setup either in MemoryMode or AppDirect mode.
With AppDirect, PM is exposed as a block device and offers data durability; we focus on the AppDirect mode of PM.

% configuration
\vheading{Interleaved or non-interleaved setup}.
With the AppDirect mode,
individual \nvdimms can be setup in interleaved or non-interleaved mode.
Non-interleaved mode exposes each \nvdimm as its own PM device.
Interleaving exposes multiple \nvdimms within a NUMA node
  as a single PM device where 4KB blocks are spread out (round-robin)
  across available \nvdimms.

% 256B block size
\vheading{Write amplification}. 
Each \nvdimm has a write combining buffer (XPBuffer) that
batches writes into 256 bytes
  before writing to the underlying media.
Writes smaller than 256 bytes force the XPBuffer to perform
  read-modify-writes which results in write amplification
  and introduce \vio overheads.

% threads
\vheading{Thread scaling}.
Using too many threads to perform writes on \nvdimms
  causes writes to queue quickly at the XPBuffer;
  XPBuffer introduces head-of-line blocking overheads as it
  combines writes and manages the load~\cite{izraelevitz2019basic,yang2020empirical,daase2021maximizing,zhou2022odinfs}.

% why mixed IO is bad
\vheading{Mixed I/O}. 
\nvdimms are connected to channels which in turn are
  connected to an integrated memory controller (iMC) on the CPU.
Read and write requests are inserted into the read or write pending queues
  inside the iMC.
Due to the asymmetric latencies of PM,
  processing reads and writes concurrently at peak bandwidth utilization
  incur blocking overheads at the iMC \cite{izraelevitz2019basic,yang2020empirical,daase2021maximizing}.

% CXL
\vheading{CXL and PM expansion}.
Compute Express Link (CXL)~\cite{cxl} is the first open multi-protocol
method to support a cache-coherent interconnect for processors
and memory devices. CXL is built upon PCIe and supports memory
expansion~\cite{jung2022hello,maruf2022tpp,shan2022towards,gouk2022direct}.
With CXL, \nvdimms can also be hosted and
  accessed over the PCIe-bus like SSDs.
CXL allows CPU cacheable loads and stores to PM over the PCIe-bus.
Recent studies on real CXL hardware show that
  like PM, CXL performance also suffers with too many threads, is sensitive
  to access patterns and how data is distributed to CXL devices~\cite{sun2023demystifying}.
With CXL, available PM capacity and bandwidth in a server will scale up
  significantly, making PM a promising building block for
  large-scale, high-throughput, persistent key-value stores.

\subsection{Persistent Memory Key-Value Stores}
Prior PM stores
% avoid giant citation lists like this, looks bad
%~\cite{seltzer2017pmemcached,zhang2021chameleondb,kannan2018redesigning,kaiyrakhmet2019slm,scargall2020pmemkv,wang2022pacman,xia2017hikv,han2020lightkv,benson2021viper,Lersch2020,Yao2020,chen2020flatstore}
  allow application threads to
  directly access PM, and
  rely on memory controllers to place
  data across individual \nvdimms.
Thus, PM stores offload
  fine-grained control over the \vio on PM to applications and hardware.
They support direct-access for applications and 
  provide low latencies (2--5 us/op).
We analyze existing PM key-value stores and observe that
  they utilize only a fraction of PM bandwidth
  and suffer from low throughput and scalability,
  as shown in Figure-\ref{fig:scalability-others}.

We measure the peak bandwidth utilization of these PM stores
  for pure-write (YCSB LoadA) and pure-read (YCSB RunC) workloads
  using 8B keys and with varying value sizes (from 8B up to 1kB).
We use a single NUMA node with six \nvdimms and up to 48 threads.
For this study, we categorize existing PM key-value stores into two groups:
  key-value stores that are retrofitted for PM and ones that
  are designed from the ground up for PM.

% PM has low write bandwidth and the nuanced performance
%   profile of PM makes it challenging for existing
%   PM key-value stores to achieve high PM write bandwidth
%   utilization.
% We analyze existing PM key-value stores and highlight that
%   they utilize only a fraction of PM write bandwidth.
% We categorize existing PM key-value stores into two groups:
%   stores that are built ground-up for PM or
%   retrofitted for PM.

\vheading{Retrofitted PM stores}.
Retrofitted PM stores are derived from key-value stores
  that are originally designed for block devices \eg
  stores like RocksDB~\cite{rocksdb} and
  LevelDB~\cite{ghemawat2011leveldb} that are based on log-structured merge trees~\cite{o1996log}, or
  from in-memory key-value stores \eg pmem-Redis~\cite{pmemredis}.
ChameleonDB~\cite{zhang2021chameleondb}, MatrixKV~\cite{Yao2020},
  NoveLSM~\cite{kannan2018redesigning}, SLM-DB~\cite{kaiyrakhmet2019slm},
  and ListDB~\cite{kim2022listdb} are a few
  retrofitted LSM-based PM key-value stores.
These stores do not cater to the unique characteristics of PM but
  take advantage of its byte-addressability and durability.

We observe that the peak write bandwidth achieved across NoveLSM,
 SLM-DB and pmem-Redis~\cite{pmemredis} is 215 MB/s
 which is <2\% of the available PM write bandwidth.
Amongst the retrofitted stores,
    ChameleonDB obtains $\approx$29\% of available write bandwidth,
    as shown in Table-\ref{tab:tbl-mot}.
Since these PM stores are not NUMA-aware,
  their performance does not scale to multiple NUMA nodes.

\vheading{New PM stores}.
Key-value stores that are custom-built for PM like
  FlatStore~\cite{chen2020flatstore} and Viper~\cite{benson2021viper}
  achieve better write-bandwidth utilization \eg
  with larger 1kB values they obtain $<$45\%
  of available write bandwidth, as shown in Table~\ref{tab:tbl-mot}.
These PM stores assume a single interleaved PM device.
Viper customizes to PM by allowing application threads to
  write directly to PM without intermediate DRAM buffers,
  and has \nvdimm-aligned logs and evenly distributes threads
  across logs.
FlatStore proposes efficient batching to avoid small writes
  to PM and uses per-core
  logs that span across multiple \nvdimms.
Thus, FlatStore and Viper do not limit the
  number of threads accessing a single \nvdimm.
Thus, these PM stores do not maintain fine-grained control
  over the \vio on \nvdimms, and hence
  utilize a small fraction of PM bandwidth.
Further, they have low throughput that degrades with increasing number of application threads (beyond 8), as shown in Figure-\ref{fig:scalability-others}.

% \usepackage{booktabs}
% \usepackage{graphicx}
%\begin{table}[]
%\centering
%\begin{tabular}{@{}ccc@{}}
%\toprule
%PM key-value store  & Peak bandwidth   & \% utilization \\ \midrule
%ChameleonDB         &    3.7 GB/s      &        29\%            \\
%NoveLSM             & 215 MB/s         &        2\%            \\
%pmem-Redis          & 100 MB/s         &        1\%            \\ 
%SLM-DB              & 20 MB/s          &        0.25\%         \\ \midrule
%FlatStore           &   5.5 GB/s       &       43\%         \\
%Viper               &   1.5 GB/s       &       12\%         \\ \midrule
%\end{tabular}%
%\mycaption{PM stores}{Peak PM write-bandwidth utilization
%on a single node (with 6 \nvdimms and up to 48 threads) for
%YCSB LoadA with 8B keys and variable size values (8B-1kB).}
%\label{tab:tbl-mot}
%\end{table}

%  baseline: 12.7 GB/s

\begin{table}[]
\centering
\begin{tabular}{@{}ccc@{}}
\toprule
PM key-value stores & \begin{tabular}[c]{@{}c@{}}Write bandwidth\\ (\% utilization)\end{tabular} & \begin{tabular}[c]{@{}c@{}}Read bandwidth\\ (\% utilization)\end{tabular} \\ \midrule
ChameleonDB  & 4 GB/s (29\%)                                                               & 2 GB/s (7\%)                                                               \\ \midrule
FlatStore    & 6 GB/s (44\%)                                                               & 3 GB/s (10\%)                                                               \\
Viper        & 2 GB/s (12\%)                                                               & 3 GB/s (9\%)                                                               \\ \bottomrule
\end{tabular}
\mycaption{PM stores}{Peak PM bandwidth utilization
of PM stores on a single node (with 6 \nvdimms, up to 48 threads) for writes and reads with 8B keys and across 8B--1kB values.}
\label{tab:tbl-mot}
\if\removespace 1
\vspace{-25pt}
\fi
\end{table}

%\begin{table}[]
%    \resizebox{\columnwidth}{!}{%
%        \begin{tabular}{@{}ccccc@{}}
%            \toprule
%            PM stores &
%              \multicolumn{2}{c}{\begin{tabular}[c]{@{}c@{}}Write bandwidth\\ (\% utilization)\end{tabular}} &
%                  \multicolumn{2}{c}{\begin{tabular}[c]{@{}c@{}}Read bandwidth\\ (\% utilization)\end{tabular}} \\ \midrule
%                        ChameleonDB & 3.7 GB/s & 29.1\% & 2.1 GB/s & 6.5\% \\ \midrule
%                        FlatStore   & 5.5 GB/s & 43.6\% & 3.0 GB/s & 9.6\%   \\
%                        Viper       & 1.5 GB/s & 12.2\% & 2.9 GB/s & 9.3\%   \\ \bottomrule
%        \end{tabular}%
%        }
%\mycaption{PM stores}{Peak PM write and read bandwidth utilization
%on a single node (with 6 \nvdimms and up to 48 threads) for
%YCSB LoadA and RunC with 8B keys and variable size values (8B-1kB).}
%\label{tab:tbl-mot}
%\end{table}

\vheading{Direct access}.
Current PM stores provide low latency for applications and 
  allow application threads to directly access PM.
PM stores no longer control the number of
  concurrent threads accessing an \nvdimm; they offload this control to applications.
Next, PM stores rely on hardware to combine multiple \nvdimms
  into a single interleaved PM device.
They let go of fine-grained control over how data is placed across
  individual \nvdimms; a large file round-robins at 4kB granularity across available \nvdimms in a NUMA node.
Thus, PM stores do not control how data is placed across
  individual \nvdimms.
Since PM stores trade off fine-grained control over PM
  for low latency, they obtain a fraction of PM bandwidth and suffer from low throughput and scalability.

\vheading{Running a key-value store on a PM file system}.
One might wonder if running a traditional key-value store on a file system designed for PM results in good performance.
We ran RocksDB on OdinFS, a PM-customized POSIX file system
  which aims to achieve high performance and scalability by
  retaining fine-grained control over PM.
We find that RocksDB does improve by using OdinFS (up to 35\%),
but it is still significantly lower than the performance of PM key-value stores such as FlatStore. 
One reason behind this is that OdinFS does not support the \texttt{mmap} operation;
most PM key-value stores access PM via memory mapping, since accessing it via read and write systems calls can be 6--16\myx slower.
More fundamentally, even if OdinFS extends support for mmap,
  it cannot provide hugepages (as it relies on stripping PM across NUMA nodes for thread parallelism and scalability);
  hugepages directly impact the
  performance of \mmap applications~\cite{kadekodi2021winefs}.
Thus, one cannot obtain a write-optimized PM store by running a key-value store built for solid state drives on top of a PM file system. 

\if 0
OdinFS~\cite{zhou2022odinfs} is PM-customized POSIX file system
  which aims to achieve high performance and scalability by
  retaining fine-grained control over PM.
Traditional system-call based key-value stores
  like RocksDB~\cite{rocksdb} can benefit
  from running on OdinFS instead of on PM-unaware file systems.
We observe that with a single-threaded client,
  RocksDB on OdinFS has 22\% higher performance on a single node
  and 35\% higher performance on four nodes relative to
  RocksDB on ext4~\cite{ext4fs} (OdinFS stalls or crashes
  with multiple clients).
However, PM stores memory-map and perform direct loads and stores
  to PM (that is the most efficient way of accessing PM).
For instance, a single thread performing
  read and write system calls is 6\myx and 16\myx slower
  than a thread using loads and stores respectively.
All state-of-the-art PM stores (Table~\ref{tab:tbl-mot})
  use the \mmap interface and bypass the underlying file system.
Although OdinFS designs for indirect-access to application,
 memory-mapping applications cannot benefit from its
 architecture.
We observe that currently OdinFS does not support mmap interface.
More fundamentally, even if OdinFS extends support for mmap,
  it cannot provide hugepages (as it relies on stripping PM across NUMA nodes for thread parallelism and scalability);
  hugepages directly impact the
  performance of \mmap applications~\cite{kadekodi2021winefs}.
By merely running an existing PM key-value store on top of
  OdinFS, we cannot achieve high performance or scalability.
\fi

\vheading{Summary}. 
We need a new PM store that 
  takes advantage of the high and expandable
  bandwidth of PM and provides scalable write throughput
  to applications.
Such a high-throughput PM store must be
  tailored to the nuanced performance of PM
  and must account for the best practices that enable
  high PM bandwidth utilization.

\section{Analysis}
\label{sec:Analysis}

We perform an empirical study to understand how to obtain high PM
  bandwidth utilization and summarize our analysis and prior
  work~\cite{daase2021maximizing,yang2020empirical,izraelevitz2019basic}
  into four design recommendations for PM stores.

% To understand how to obtain high PM bandwidth utilization,
%   we perform an empirical study
%   and summarize our analysis and prior work~\cite{daase2021maximizing,yang2020empirical,izraelevitz2019basic}
%   into four design recommendations for PM stores.

\subsection{Experiments}

\vheading{Setup}. We use a four-socket machine
  with 3 TB of Intel Optane DC Persistent Memory
  across 24 \nvdimms, 224 cores, 790 GB DRAM,
  and with Ubuntu 20.04 and Linux 5.16 kernel for this study. 
Our experiments use DAX-enabled ext4 to memory-map 12 GB sized
  PM files and use direct load and non-temporal store instructions~\cite{rudoff2017persistent}
  to read and write to PM.
Memory-mapping PM avoids
  \vio overheads from the underlying file system
 ~\cite{xu2016nova,kadekodi2019splitfs,kadekodi2021winefs}.
  We page fault and zero-out PM pages before using them to avoid
    their overheads in our experiments.

In our experiments, we perform reads and writes to PM in
  different configurations (interleaving vs non-interleaving modes, single vs multiple NUMA nodes),
  with varying I/O sizes, I/O patterns (sequential vs random), and the
  number of reader and writer threads.
We make the following observations.

\vheading{O1: Sequential access gives better performance than random}.
With random instead of sequential 512B accesses,
single-threaded read throughput drops by $\approx$2\myx
  from 3.9 GB/s to 2.1 GB/s, and
  write throughput drops
  by $\approx$3\myx from 2.2 GB/s to 0.7 GB/s.
Thus, sequential access pattern is better for reads and writes;
sequentiality is more important for writes than reads.

\vheading{O2: 512B access size gives the best performance}.
We note a peak single-threaded throughput of $\approx$2.2 GB/s
  with 512B and 1kB sized writes,
  and $\approx$3.9 GB/s with 256B and 512B sized reads;
  512B is preferable for both reads and writes.

\vheading{O3: Maximum of 3 writers and 6 readers per \nvdimm}.
On a single \nvdimm with 3 writers, we observe a
  sequential write throughput of 2.2 GB/s; 
  with 6 readers, we observe a
  sequential read throughput of 6.9 GB/s. 
However, increasing the number of threads
  decreases the write and read throughput
  due to head-of-line blocking overheads.
However, we observe random write throughput of 0.6 GB/s with 3 writers
  and random read throughput of 5.85 GB/s with 6 readers;
  random reads scale better than random writes.
% vc: unclear what this means?
%Further, with increasing ratio of DRAM-to-PM accesses,
%  the number of concurrent threads required to obtain
%  high bandwidth utilization increases; reads require more threads than writes.

\vheading{O4: Mixed \vio on \nvdimms}.
Contrary to prior PM analysis~\cite{daase2021maximizing},
 we observe that performing concurrent reads and writes to PM yields better throughput.
Concurrent reads and writes to PM obtain
 higher cumulative bandwidth although they reduce the peak
 read and write PM bandwidth utilization.
With 2 readers and 1 writer, we observe a peak \vio throughput of 3.31 GB/s per \nvdimm with mixed \vio.
However, serializing reads and writes to \nvdimms, requires
  8 readers and 1 writer to read at 6.8 GB/s and write at 2.14 GB/s
  but achieves a cumulative throughput of $\approx$3.26 GB/s.
At peak PM write and random-read bandwidth utilization, avoiding mixed \vio provides $<$5\% higher throughput.
Thus, concurrent \vio uses 3\myx fewer threads and allows higher throughput in most cases.

% With 4 readers and 1 writer, we observe 2.83 GB/s read throughput
%   (41\% of read bandwidth),
%   and 1.67 GB/s write throughput (76\% of write bandwidth).
% When operating at peak bandwidths,
%   overall throughput also drops with mixed \vio.
% With 16 readers and 4 writers, avoiding
%   mixed \vio yields 5\% better throughput.
% However, with lesser threads (when bandwidth is underutilized)
%   performing concurrent \vio yields better throughput.

\vheading{O5: Non-interleaved \nvdimms provide better performance}.
We observe consistent and higher bandwidth utilization
  with the non-interleaved mode.
For example, with a single thread, interleaving \nvdimms has
  up to 50\% lower throughput for reads and writes across
  varying value sizes (512B--4kB writes).
Even with a higher number of threads, managing individual non-interleaved \nvdimms yields up to 20\% higher throughput relative to interleaving.
With interleaving and write sizes ranging from 256B to 2MB,
  write throughput drops from 12.7 GB/s to 10.5 GB/s
  ($\approx$85\% of the maximum write bandwidth), while
  it remains consistently above 11.7 GB/s ($\approx$97\% utilization) without interleaving.
% We also observe a maximum read and write throughput
%   of 6.9 GB/s and 2.2 GB/s per non-interleaved \nvdimm,
%   and performance scales to multiple \nvdimms.
% On a single NUMA node with 6 non-interleaved \nvdimms,
%   we observe sequential read and write throughputs of
%   41.4 GB/s and 12.7 GB/s (\ie 6.8 GB/s and 2.1 GB/s per \nvdimm)
%   respectively.
% However, interleaving \nvdimms forces concurrent reads and writes
%   to PM; with a 1:1 read-to-write ratio and 512B values,
%   write bandwidth drops to 10.1 GB/s
%   ($\approx$79\% of the maximum)
%   and read bandwidth drops to 9.20 GB/s
%   (22\% of the maximum).

\vheading{O6: Avoid cross-node traffic}.
We observe 3\myx lower write throughput,
  and 2.5\myx lower read throughput if threads accessing \nvdimms
  are not pinned to their local NUMA node; cross-node traffic
  limits throughput and scalability.

\vheading{Summary}.
With 4 nodes and 24 \nvdimms
  we observe a peak aggregate throughput of 162 GB/s for sequential reads (96\% of the theoretical limit), 124.2 GB/s for random reads,
  and 48.3 GB/s for sequential writes (92\% of the theoretical limit);
  we estimate theoretical limits by assuming perfect \nvdimm and NUMA scalability.
On a single \nvdimm, we also observe a maximum random read and sequential write throughput of 5.85 GB/s and 2.2 GB/s.
With 6 \nvdimms on a single node, we observe a
  peak PM throughput of 12.7 GB/s for sequential writes and
  31.5 GB/s for random reads.
Note that we use these as our baselines for the maximum attainable
    read and write bandwidth in the rest of the paper.

%%%%%%%%%%%%%%%%%
% vc: dont think we need this, comes off as weird
\if 0

\viheading{Write-optimization}.
Our analysis highlights that with random access patterns,
  write throughput suffers more than read throughput.
With random accesses and multiple threads,
  write throughput suffers by $\approx$5\myx while
  random reads achieve 80\% of sequential read performance.
Thus, this paper proposes a write-optimized PM store.
% Cite here!
\skl{This paragraph suddenly pop up. Why don't we add this to our observation (O7)?:}
\souji{Its not an observation about PM; its
our understanding.. so I thought its better here after summary}
\fi
%%%%%%%%%%%%%%%%

\subsection{Design Recommendations}
\label{sec:recommendations}

We summarize four recommendations (from our analysis and prior work) that enable high PM bandwidth utilization.

\vheading{R1. Avoid interleaved PM}.
Managing individual non-interleaved \nvdimms
  provides high throughput which remains stable
  across varying value sizes and scales with the
  available \nvdimms.
However, existing PM key-value stores (\sref{sec:Background}) and
  PM file systems~\cite{zhou2022odinfs,xu2016nova, kadekodi2019splitfs,kadekodi2021winefs}
  use the interleaving approach for simplicity
  and suffer from low throughput and scalability.
  %they need fundamental redesign to operate without interleaving.

\vheading{R2. Limit concurrent PM access}.
Increasing the number of reader or writer threads without bounds
  causes the buffers in \nvdimms to fill quickly and
  induces head-of-line blocking overheads.
Thus, it is critical to limit the maximum number of concurrent
  threads on each individual \nvdimm.

% \vheading{R3. Do not always mix read and write I/O}.
\vheading{R3. Mix read and write \vio}.
Performing concurrent reads and writes provides better throughput
  and requires fewer threads to achieve peak bandwidth
  utilization (3.31 GB/s per \nvdimm).

% vc: what does this line mean???
%We observe that with low PM capacity, saturating PM \vio bandwidth 
%  requires fewer threads per \nvdimm than with high PM capacity (multiple \nvdimms across NUMA nodes).
%Thus, mixed \vio reduces CPU utilization
%  and provides better throughput.

\vheading{R4. Limit cross-node traffic}.
Memory and thread sharing across multiple NUMA nodes results in
  cross-node traffic and introduces overheads from the
  high-latency low-bandwidth NUMA interconnect.
Thus, minimizing cross-node traffic is crucial for high
  PM bandwidth utilization and scalable throughput.

\section{\sysname: Design}
\label{sec:design}

We present \sysname, a write-optimized key-value store for persistent memory (PM).
%\sysname supports increasing number of application threads
 % and provides high throughput that scales to multiple NUMA nodes.
\sysname provides a simple interface with \texttt{put}, \texttt{get}, and \texttt{delete} operations.
\sysname supports strong consistency (linearizable reads and writes).
\sysname maintains crash consistency in the event of a crash and recovers efficiently.

The key insight powering \sysname is that
  PM stores must maintain \emph{fine-grained control over all PM accesses}
  to achieve high and scalable throughput.
This idea follows observations from prior PM studies and the
  recommendations from our analysis (\sref{sec:Analysis}).
Existing PM stores offload the control
over PM \vio to hardware and applications.
They rely on memory-controllers to distribute data across individual
  non-volatile DIMMs (\nvdimms) and allow application threads to directly
  access the data on PM.
Thus, existing PM stores harvest the low latency of PM.
However, this fundamentally trades off high throughput as PM stores no longer control
  how data is placed on individual \nvdimms (violates R1), and how many threads
  access an \nvdimm concurrently (violates R2).

\sysname is the first PM store to have
  fine-grained control over all PM accesses and to
  leverage the high bandwidth of PM.
It proposes an architecture with \emph{indirect-access to PM for applications}.
\sysname uses three main components to implement this architecture
  and follows the four design recommendations from our analysis.

\vheading{Log interface for \nvdimms}.
Instead of using a single hardware-managed PM device, \sysname
  manages individual \nvdimms and implements a log abstraction on top of each \nvdimm.
By writing to a \nvlog provided log, \sysname writes sequentially to \nvdimms and makes data placement decisions at the granularity of each put request (R1).

\vheading{Dedicated worker threads}.
Instead of allowing applications to access the data on PM directly,
  \sysname uses dedicated workers that perform \vio on behalf of applications.
\sysname ensures that a fixed number of workers perform \vio on a single \nvdimm (R2).

\vheading{Leveraging other media}.
With \nvlog and \workers, \sysname efficiently 
process write requests.
However, to enable efficient reads,
  \sysname leverages DRAM and disks.
\sysname maintains in-memory indexes and
  checkpoints periodically to disks.
Thus, \sysname avoids overloading PM and obtains
  good read and write throughput simultaneously.

\vheading{NUMA awareness}.
\sysname adopts a NUMA-aware design.
\sysname partitions data and metadata
  across NUMA nodes
  and uses workers from the same node to access the data. This enables scalable throughput across multiple NUMA nodes (R4).

%%%%%%
%vc: doesn't add much
\if 0 
With \nvlog at the center,
  \sysname is designed to write
  sequentially to \nvdimms and is write-optimized.
Further, designing for high bandwidth utilization
  and by leveraging DRAM and disks,
  \sysname aims to simultaneously achieve
  good read and write throughput.
We now present the high-level architecture of \sysname.
\fi 
%%%%%%%

% With \nvlog at the heart of the design,
%  \sysname is a write-optimized key-value store 
% Note that designing only for efficient writes can have prohibitively
%   low read throughput and high CPU utilization.
% \sysname achieves the following goals simultaneously.
% \sysname obtains high PM write-bandwidth utilization,
%   has good read throughput and low CPU utilization.
% We now discuss the architecture of \sysname
%   and describe its design in detail.

\subsection{Architecture: Indirect-Access to PM}
\begin{figure}
  \centering
\includegraphics[width=0.95\columnwidth]{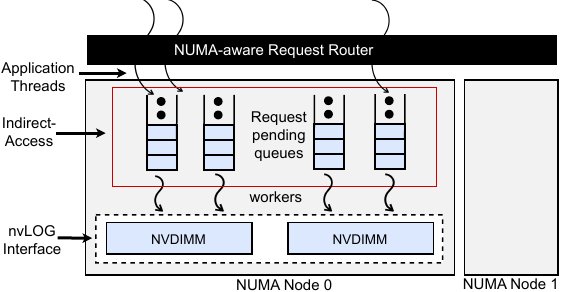}
\if\removespace 1
\vspace{-10pt}
\fi
\mycaption{Architecture}{
    \sysname uses a log interface to \nvdimms (\nvlog)
    and provides indirect-access for applications. 
    \sysname uses dedicated workers
    to process requests on behalf of applications and \sysname is NUMA-aware.}
  % \sysname builds on \spm,
  % leverages multiple media in its partitioned data plane, and
  % employs an \vio-scheduler in its shared control plane.}
\label{fig:overview}
\end{figure}

% vc: This seems like the same stuff as in section 4 start

\sysname is designed for machines with multiple 
  non-uniform memory access (NUMA) nodes and
  multiple \nvdimms per node.
\sysname manages individual non-interleaved \nvdimms
  and provides indirect-access for applications.
Figure~\ref{fig:overview} illustrates the indirect-access
  architecture of \sysname.

%\noindent \vspace{2pt}
First, \sysname exposes a log interface on top of individual \nvdimms (\nvlog). 
Each \nvlog log tailors \vio to the nuanced performance profile of PM.

Next, \sysname uses dedicated workers that read and write
  to \nvdimms using \nvlog.
\sysname ensures that the number of concurrent
  threads accessing a single \nvdimm does not exceed a
  threshold.
Each worker has its own read and write pending queues
  and processes the requests in those queues.
  
Application threads can enqueue requests into
  a worker's request pending queue and wait for their completion.
\sysname employs a NUMA-aware router that 
  directs the application threads to a specific worker's
  request pending queues.
This indirect-access architecture allows \sysname to reclaim
  the fine-grained control from hardware and applications,
% This fine-grained control is
  which is
  crucial for achieving high and scalable throughput.

%%%%%
% vc: not needed

\if 0
In this section,
  we describe \nvlog: the log interface to \nvdimms,
  indirect-access for applications,
  and outline how \sysname simultaneously achieves
  good read throughput alongside
  NUMA-scalable write throughput.
Then, we 
summarizes the life of a read and write request in \sysname,
and outline how \sysname extends to other PM media.
% add section labels.
\fi
%%%%%

\subsection{Log Interface to \nvdimms}

\sysname manages PM as multiple, individual,
  non-interleaved PM devices and implements a 
  log interface for \nvdimms (\nvlog).

% Interface
\nvlog presents a high-level interface on top of \nvdimms.
\nvlog allows \sysname to
  create, read, write, and delete
  logs (append-only files) on individual \nvdimms
  without worrying about low-level details
  like the complex performance characteristics of PM.

Traditionally, PM key-value stores have used interleaved PM.
Interleaving combines multiple \nvdimms into one PM device;
  sequential PM regions map across \nvdimms in round-robin
  fashion at 4kB granularity.
A file system is mounted on top of the interleaved PM device.
PM stores \mmap files and store their data and metadata on PM.
While using a single interleaved PM device per NUMA node is convenient,
  it gives up fine-grained control over the \vio at each \nvdimm,
  which is crucial for obtaining high throughput (\sref{sec:Analysis}).

\nvlog replaces this stack (interleaved PM and file system).
Instead of a file abstraction, \nvlog exposes the log abstraction (reads and appends which translate to random reads and sequential writes to \nvdimms).
\nvlog is tailored for the performance characteristics of PM and the needs of \sysname.
Instead of one interleaved PM device, \nvlog manages multiple
  \nvdimms each as its own PM device.
\nvlog mounts a file system on each \nvdimm that it manages.
\nvlog pre-allocates large files, zeroes-out and pre-faults them
  (so that page faults and the overheads from zeroing pages are
  not incurred in the critical path of writing to the logs).
As a result, obtaining a new log from \nvlog is inexpensive.

% Batching
\nvlog also implements efficient log reads and appends.
For instance, \nvlog uses non-temporal stores~\cite{rudoff2017persistent}
  and AVX-enabled memcopies~\cite{avx}
  as they enable high PM bandwidth utilization.
\nvlog is aware of the block size of PM, so
  \nvlog batches log appends and writes to the media at 256 bytes
  or larger granularity.
This prevents write amplification and avoids PM \vio bottlenecks.

\subsection{Workers and Request Queues}

%\sysname maintains data and metadata logs on \nvdimms using \nvlog.
\sysname maintains a fixed number of workers per \nvdimm.
\sysname associates each data and metadata log with a worker.

Worker threads process requests on behalf of application threads.
Application threads talk to a centralized router, which points them to a worker.
The application thread then directly enqueues its request in the worker's read or write pending queue.
To process writes, workers dequeue a write request,
  write values to the data log and corresponding metadata
  to the metadata log; \nvlog batches small writes to \nvdimms.
To process reads, workers dequeue a read request,
  and read its value from data logs.

\sysname couples data storage with data accesses (by associating a set of
  data and metadata logs to one worker).
\sysname decouples PM accesses from application threads
  using dedicated workers and request queues.
This may seem counter-intuitive as one of the advantages of PM is that
  applications can directly read or write to PM at low latency.
However, to achieve high throughput and leverage the high bandwidth of PM,
  PM stores must carefully control \vio to PM; \sysname maintains its own threads
  instead of allowing applications to access the data on PM.
With indirect-accesses, \sysname achieves high throughput
  that scales with increasing number of application threads.

\vheading{NUMA awareness}.
\sysname associates workers with a single data and metadata log
  on a particular \nvdimm.
Workers always access the \nvdimms in their own NUMA node.

\vheading{CPU utilization}.
Using dedicated workers increases the 
  CPU utilization of \sysname.
Prior PM studies recommend
avoiding
  concurrent reads and writes on a single \nvdimm to 
  ensure high write-bandwidth utilization;
however, we observe that concurrent read and write \vio provides higher throughput
  with fewer number of workers.
We use this in the design 
  to lower the CPU utilization of \sysname.
\sysname allows multiple workers with data and metadata logs on the same \nvdimm to process read and write requests simultaneously.
% We can talk about our experience with building phased-IO
% Seeing that it has 10% lower throughput and requires 2x more workers

\subsection{Request Routing}

\sysname has a centralized request router that is shared across NUMA nodes.
The router in \sysname has two main functions:
  serializing requests, and routing application threads to workers.

\vheading{Serializing requests}. 
The router is the serialization point in \sysname. 
All put requests are assigned a serial number that is persisted as part of the metadata,
and get requests are assigned metadata.
% what metadata?

\vheading{Routing}.
For put requests, the router distributes application threads
uniformly across the available workers.
The router ensures that the write request load is distributed 
  uniformly across NUMA nodes and \nvdimms.
For get requests, application threads are routed to a specific worker that manages the corresponding data log.

\vheading{Read throughput}.
\sysname partitions data and metadata across NUMA nodes
  and \nvdimms.
Thus, the router in \sysname must lookup the latest metadata to locate the data log with the latest data to direct
  application threads to that worker's request queues.
\sysname returns the metadata of a write request once
  it is processed and allows applications to cache this metadata.
Thus, application threads can pre-populate get requests with
  metadata and can enqueue the requests into the worker's queue; in this fast-path, router does not have to perform
  metadata lookups while routing the get request.
Request routing is performed outside the throughput-critical path of \sysname.

% Further, the router uses the metadata (merge logs and the index) from all NUMA nodes
% to locate the most recent metadata. 
% Once the right \nvdimm and the data log
%   with the most recent data is identified,
%   the get request is routed to the corresponding data layer.
%   \rohan{The next statement is vague, how to get to the fast path? Does \sysname give the metadata info to the applications on get requests, so subsequent get requests have metadata? Or is there some other way? Also, the next statement is not complete, not sure how to complete it.}
% In the fast path \ie when get requests have the routing metadata
%   along with the key, router distributed the 

\subsection{Leveraging DRAM and Disks}

\sysname maintains in-memory indexes that store metadata
  and help locate the data in \nvlogs efficiently.

We observe that PM indexes~\cite{lee2019recipe,Nam2019,Zuo2018,Ma2021,chen2021hashindexes,Kim2021,Wang2021,Chen2020,Zhou2019,Lu2020}
  provide low throughput ($<$10 Mops/s) as they do not
  account for the \vio bottlenecks from inefficient PM
  accesses.
For example, PM hash tables  perform
  small, random, mixed \vio on PM,
  resulting in low overall throughput.
Thus, \sysname employs DRAM indexes.
  These indexes offer considerably higher throughput relative
  to PM indexes and do not overload PM.

However, updating in-memory indexes in the critical path of processing put requests
  leads to poor performance.
Instead, \sysname writes index updates to an in-memory \emph{merge log}
% \rohan{is this append-only? -- yes}
  in the critical path.
Using these merge logs, indexes are updated in the background.
Merge logs must be examined to fetch the most recent metadata while routing get requests;
  \sysname uses bloom filters to lower this cost.
With this design choice,
  \sysname incurs additional overheads (merge log lookups)
  while routing get requests and achieves high throughput
  for puts (avoids index updates in the critical path).
% \sysname explicitly tradesoff read performance in exchange for
%   higher PM write-bandwidth utilization.

\vheading{Checkpoints}.
Since the data layer also persists metadata using \nvlog,
  \sysname can use metadata logs
  to recover from a power failure or a crash.
However, recovering purely from metadata logs requires
  reading complete logs.
To avoid reading full logs and reconstructing entire indexes
  during recovery, \sysname checkpoints indexes.
To prevent checkpointing traffic from overloading \nvdimms,
  \sysname leverages disks and periodically checkpoints to disks.

\subsection{Life of a request}

We discuss the life cycle of a put and get request in \sysname.
Figure-\ref{fig:read-write-path} illustrates the life of a put (in red) and a get request (in green).

\begin{figure}
  \begin{center}
    {\includegraphics[width=\columnwidth]{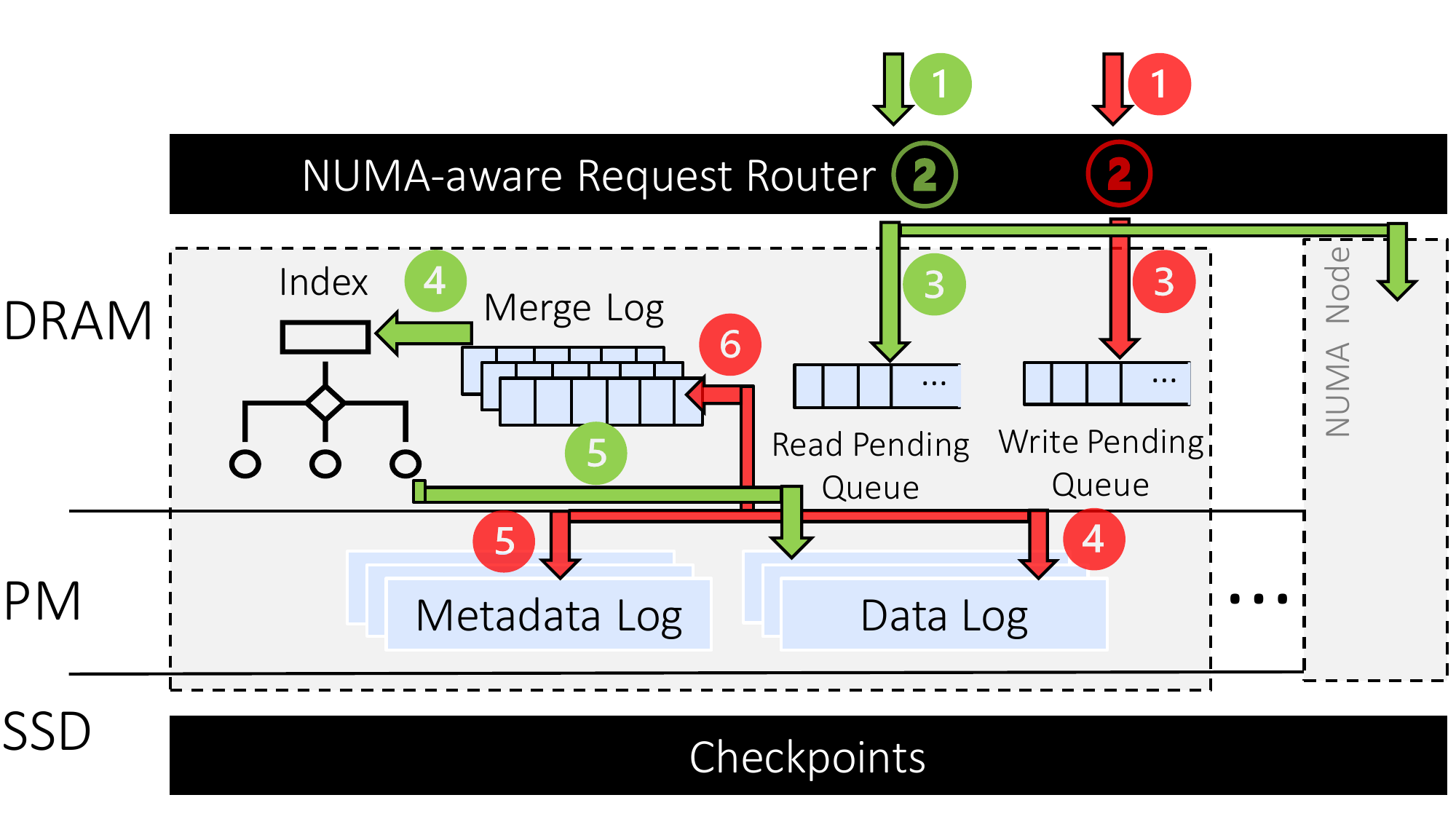}}
  % {\includegraphics[width=0.9\textwidth]{figures/skye-read-write-path.pdf}}
  % {\includegraphics[width=0.75\textwidth]{figures/lifepath.png}}
\if\removespace 1
\vspace{-20pt}
\fi
  \mycaption{Life of a request in \sysname}{
    This figure shows the life of a put request (in red color)
    and get request (in green).
    The worker's data structures are in blue, 
    per-node partitions in grey, and
    components shared across NUMA nodes in black.}
  \label{fig:read-write-path}
  \end{center}
\if\removespace 1
\vspace{-10pt}
\fi
  \end{figure}

\if 0
\begin{figure*}%
  \centering
  \subfigure[Write path in \sysname]{%
  \label{fig:write-path}%
  \centering
  \includegraphics[width=\columnwidth]{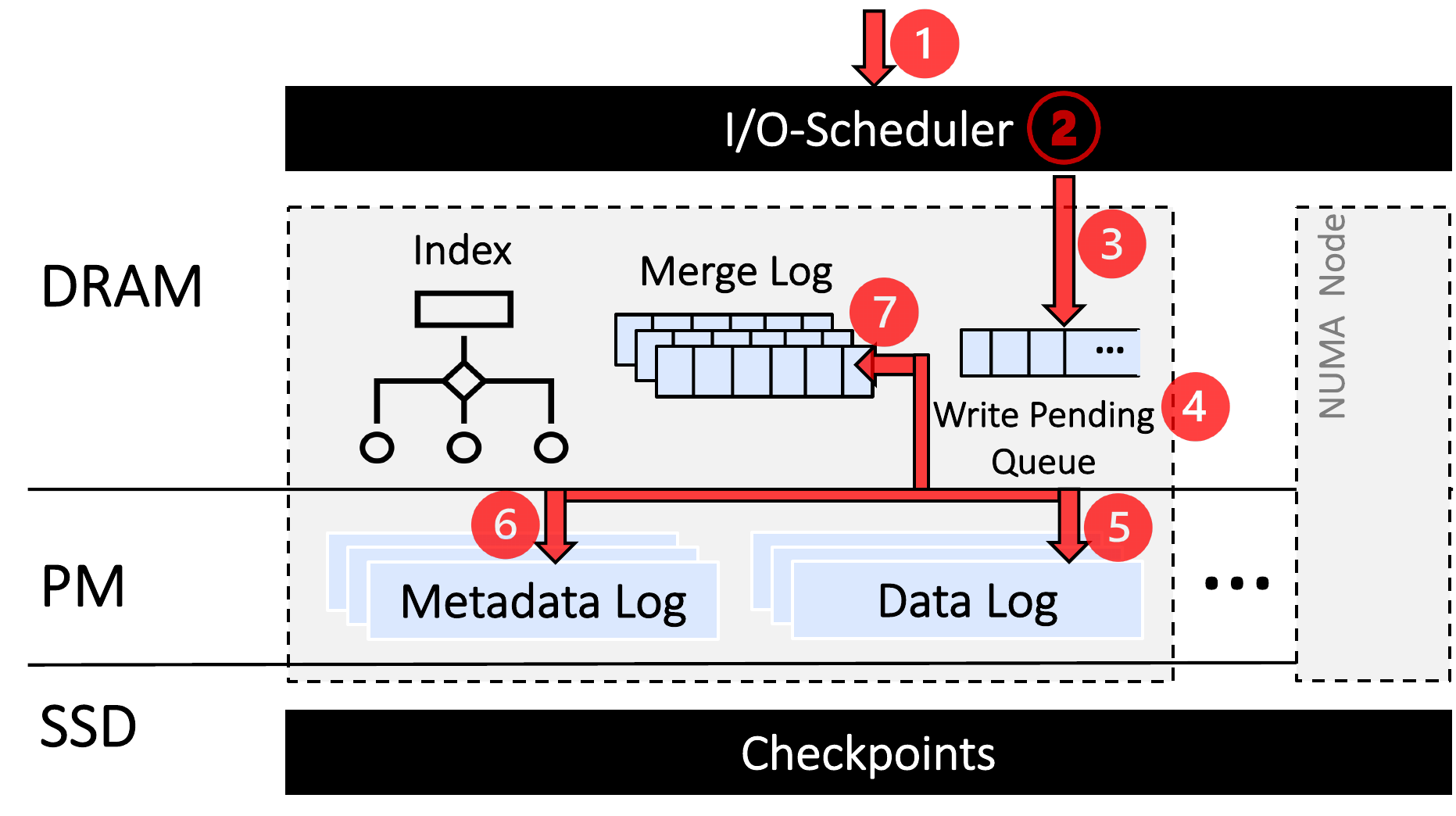}}%
  \subfigure[Read path in \sysname]{%
  \label{fig:read-path}%
  \centering
  \includegraphics[width=\columnwidth]{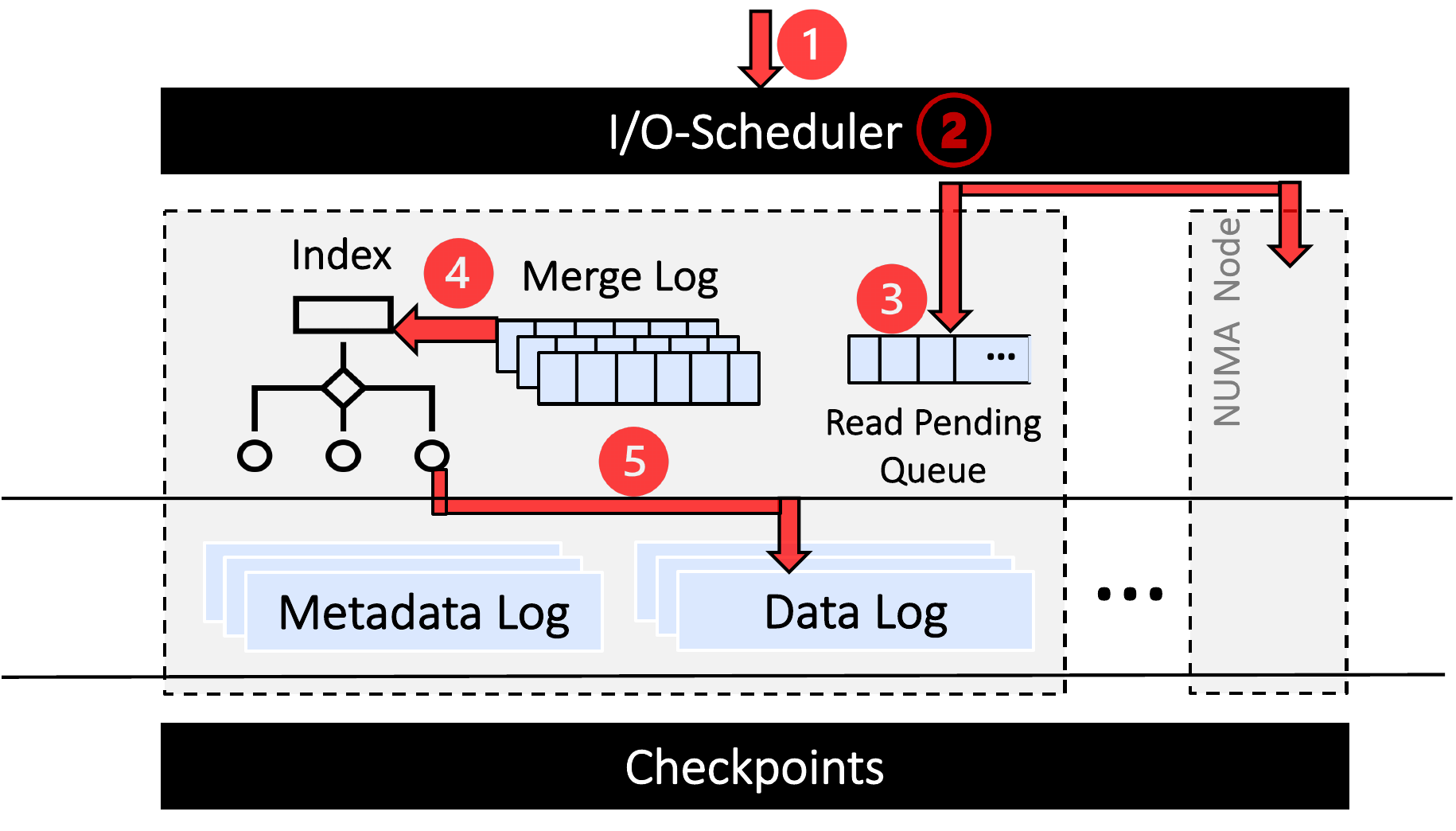}}%
  \mycaption{Life of a request in \sysname}{
    Figure-\ref{fig:write-path} shows the life of a put request
    and Figure-\ref{fig:read-path} shows that of a get request.
    Both the figures represent per-thread data structures in blue, 
    per-node partitions of the data plane in grey boxes, and
    use black to show components that are shared across
    NUMA nodes.}
  \label{fig:read-write-path}
\end{figure*}

\begin{figure}
\begin{center}
  {\includegraphics[width=\columnwidth]{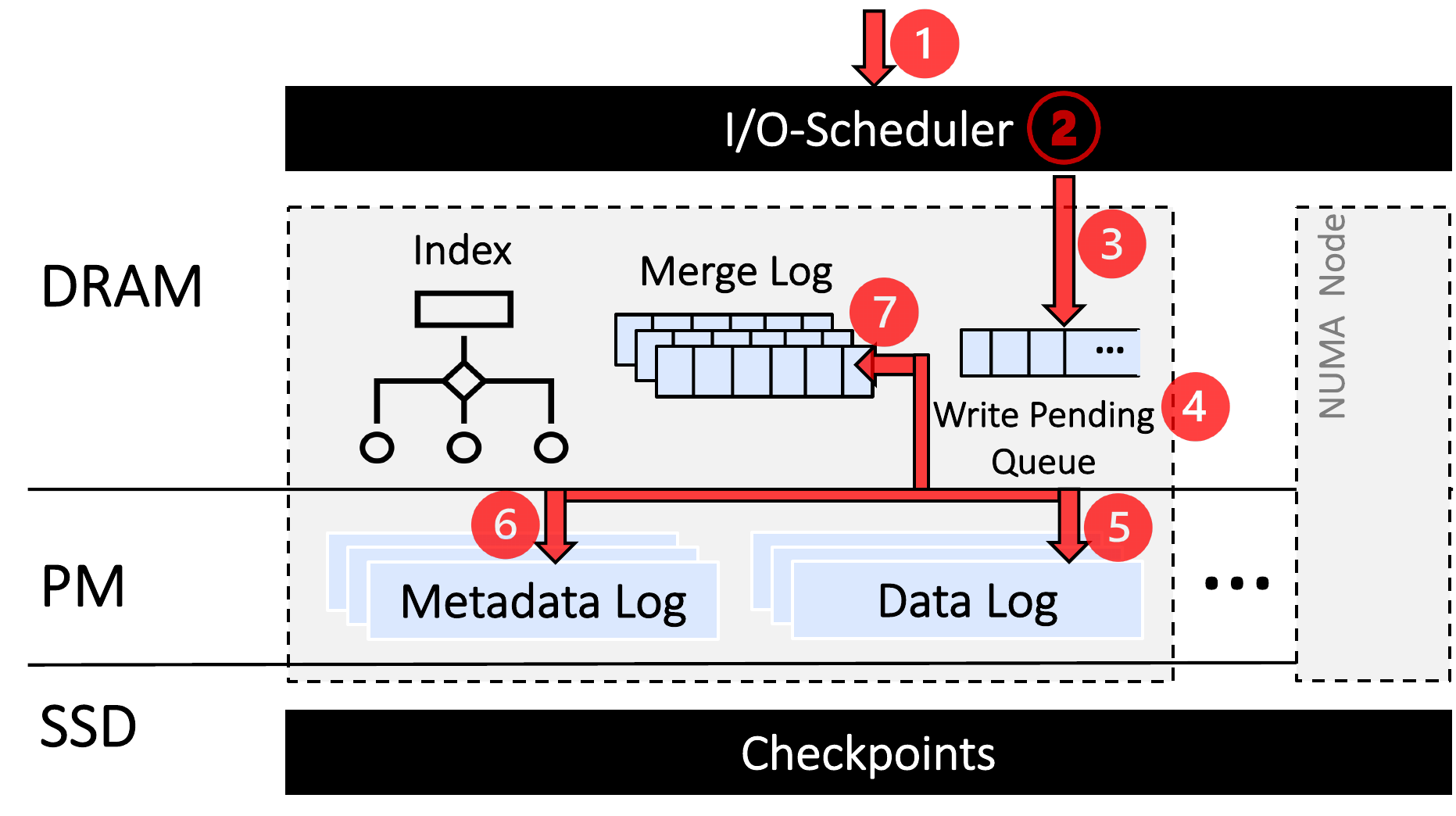}}
% {\includegraphics[width=0.9\textwidth]{figures/skye-read-write-path.pdf}}
% {\includegraphics[width=0.75\textwidth]{figures/lifepath.png}}
\mycaption{Life of a request}{
  This figure shows the life of a get request (red arrows) and put request (black arrows) in \sysname;
  per-writer data structures are in green, 
  per-NUMA-node structures are in yellow, and
  components in blue are shared across nodes.}
\label{fig:read-write-path}
\end{center}
\end{figure}
\fi

\vheading{Put Request}
\begin{itemize}[leftmargin=*,noitemsep]
    \item The put request arrives at the router.
    \item The router assigns a serial number to the put request 
      and routes it uniformly to workers across NUMA node.
    \item Application threads enqueue the request to the
      worker's write pending queue
    \item The worker dequeues a put request, appends value to the
      data log, and appends metadata to the metadata logs    
    \item Finally, worker
    appends an index update record to its in-memory merge log and
    marks the request complete.
    \item Under low load workers wait until a few requests are enqueued to avoid contention on the queues and to reduce CPU utilization; this enables \nvlog to batch writes to \nvdimms
\end{itemize}

In summary, in the critical path, workers dequeue requests,
  persist data and metadata to \nvdimms using \nvlog and append index
  update records to their corresponding in-memory merge logs.
A background merger thread reads the update records from the merge logs
  and asynchronously updates the index in the NUMA node.
 Note that application threads
can cache the metadata from put requests and send it with along
with get requests (fast path).

\vheading{Get Request}
\begin{itemize}[leftmargin=*,noitemsep]
    \item The get request arrives at the router.
    \item The router directs the read request to the worker
      that manages the data log with the latest value of the key
    \item In the fast path, clients provide the latest metadata the key. Otherwise, router uses the metadata from all NUMA nodes (index and merge logs),
      finds the most up-to-date metadata; application threads enqueues
      the get request to that specific data log's worker
    \item The worker dequeues the get request, reads the value
      from the data logs on \nvdimms, and marks
      the request complete
      % ; having separate data logs allows
      % workers to perform a single PM read to serve a get request
    \end{itemize}

In summary, in the critical path, workers 
  dequeue requests and perform a single PM read 
  to fetch the value from the data logs.
Note that application threads need not cache and pre-populate
  the metadata of get requests as
  the router looks up the metadata from all NUMA nodes
  to route a get request; routing is performed outside the
  (throughput) critical path of processing a get request.

\subsection{Crash Consistency}
\sysname ensures that it can recover to a consistent point after a
  crash or a power failure. 
\sysname orders updates and uses checksums to ensure crash consistency.
Once the metadata is appended to the metadata log only then a key-value
  is considered committed.

If a crash happens before the metadata write is done, the data is lost;
  the data log entry will be overwritten by the next write.
Partially written metadata entries will also be zeroed out.
The data log and metadata log appends can get re-ordered at PM devices
  as \sysname does not order data and metadata appends using fences
  (\eg \textit{sfence}~\cite{rudoff2017persistent}).
\sysname recovers consistently even in cases when metadata log is
  persisted but the data log is not persisted.
The metadata log contains a checksum of the data (value) so that
  \sysname can detect partial writes, bit flips, and other forms of
  corruption.
  
\nvlog interface batches writes and issues fences after persisting
  a batch of data and metadata entries.
Once data and metadata are persisted, a crash does not result
  in data loss. 
The in-memory metadata indexes will be reconstructed from
  checkpoints and the metadata logs. 

\subsection{Garbage Collection}
To support deletes, \sysname writes a new data entry with
  a higher serial number (provided by the router)
  indicating that the key is deleted.
\sysname garbage collects deleted keys when the key-value
  store starts up, by moving live keys to new data logs,
  and updating the metadata logs.
Garbage collection is also triggered when there are no
  free log segments available.

\subsection{PM Discussion}

A natural question that arises is the relevance of \sysname
  given that Intel is discontinuing its line of PM products.

The main contribution of this paper is
  our approach of empirically analyzing a new media,
  understanding how to effectively utilize it,
  and designing a high-throughput key-value store.  
  % write-optimized key-value store
  % with indirect-access architecture for applications.
We have demonstrated this for PM;
  PM requires fine-grained control over \vio to \nvdimms
  to achieve high write-bandwidth utilization and \sysname controls
  all PM accesses to achieve high and scalable throughput.
The indirect-access architecture of \sysname is a natural fit for  products like 
  Samsung's memory-semantic SSD~\cite{samsung_cxl_ssd}.
The memory-semantic SSD has very similar characteristics
  to Optane PM.
It is has a built-in DRAM cache (like XPBuffer in PM (\sref{sec:Background}))
  and is byte-addressable but has an underlying coarse-grained
  SSD media access.
To show that \sysname extends to newer media with minimal effort,
  we support traditional block-based SSD with a few lines of
  code changes within the \nvlog interface.
In the future, we envision \sysname supporting
  newer byte-addressable persistent media, by
  extending its support for optimal logging.
Our approach is also valid for PM expansion with CXL.
The write bandwidth
  of real CXL~\cite{sun2023demystifying} hardware
  is sensitive to \vio sizes, access patterns, and
  the number of concurrent accesses to the media.
We demonstrate that \sysname extends to CXL PM
  without any modifications (\sref{sec:eval}).

\section{Implementation}
\label{sec:Implementation}

We implement \sysname in 10k lines of C++ code.
By default,
    \nvlog maintains 1 GB-sized logs for data and metadata,
    uses non-temporal stores for log appends and
    AVX-512~\cite{avx} memcopies for log reads,
    and zero-pads appends to make them 8B-aligned.
\sysname maintains a fixed (configurable) number of workers per \nvdimm;
  each worker has its own
  read and write pending queues (RPQs);
  we use lock-free concurrent queues~\cite{libboost}.
Each worker maintains its own RPQs
  to avoid concurrency overheads in the critical path.
The workers in \sysname process the pending requests in their RPQs and
  go back to sleep until the application threads notify the worker.
Under low load, if request batching is enabled, the worker waits until
  a fixed (configurable) number of write requests arrive in its RPQs before
  it starts processing writes.
Batching enables high throughput and better PM write bandwidth utilization.
Workers avoid polling on the request queues to avoid contention on the queues, instead they get notified after requests are enqueued in their RPQs.
We observe that RPQs have poor scalability with increasing number of application threads and we need scalable, high-throughput RPQs that are
lock-free and wait-free to realize the benefits of indirect-access in practice.
\sysname uses FASTER~\cite{chandramouli2018faster} as its metadata index,
  implements in-memory merge logs to avoid index updates in critical path,
  and uses bloom filters~\cite{libbloom} to optimize lookups on merge logs.

\section{Evaluation}
\label{sec:eval}

We answer the following questions in our evaluation:

% \begin{itemize}[noitemsep,leftmargin=*]
%     \item What is the performance impact of $\#$workers in \sysname? (\sref{sec-eval:tunable})
%     \item How does scaling application threads impact \sysname performance, and
%     what is the throughput and latency tradeoff in \sysname? (\sref{sec-eval:put-get-latency})
%     \item What is the throughput of \sysname on a single \nvdimm?
%     Does \sysname scale to multiple \nvdimms across NUMA nodes? (\sref{sec-eval:microbenchmarks})
%     \item How does \sysname perform on workloads resembling real-world
%     applications relative to the state-of-the-art PM stores? (\sref{sec-eval:macro})
%   \item What are the overheads and tradeoffs in \sysname? (\sref{eval:overheads})
% \end{itemize}

\begin{itemize}[noitemsep,leftmargin=*]
    \item What is the throughput of \sysname? Does it scale to multiple \nvdimms and NUMA nodes? (\sref{sec-eval:microbenchmarks})
    \item What is the latency for \sysname requests? (\sref{sec-eval:put-get-latency})
    \item How does \sysname compare to state-of-the-art PM stores on workloads resembling real-world applications? (\sref{sec-eval:macro})
    \item What are the overheads and tradeoffs in \sysname? (\sref{eval:overheads})
    \item How do we tune the number of workers in \sysname? (\sref{sec-eval:tunable})
\end{itemize}

\vheading{Experimental setup}.
We use a four-socket machine with 224 cores,
  790 GB DRAM, and 3 TB of Intel Optane DC Persistent Memory,
  with Ubuntu 20.04 and Linux 5.16 kernel for our experiments.
This machine has six 128 GB \nvdimms per node,
  has Intel(R) Xeon(R) Platinum 8276 processor.
To emulate CXL PM, we use a dual-socket machine
  with 16 cores, 125 GB DRAM,
  Intel(R) Xeon(R) Silver 4314 processor, and
  with 1 TB PM (four 128 GB \nvdimms per node).
To emulate CXL PM, we pin threads to one node and access \nvdimms
  on the remote node (\sref{sec-eval:cxl}).

\vheading{Comparison points}.
We compare \sysname against three state-of-the-art PM stores:
  FlatStore~\cite{chen2020flatstore},
  Viper~\cite{benson2021viper}, and 
  ChameleonDB~\cite{zhang2021chameleondb}.
Since FlatStore and ChameleonDB are not publicly available, we use
  their implementations from the authors of Pacman~\cite{wang2022pacman}.
  We use FlatStore-H (with a hashtable index) implementation and
  observe that the measured performance aligns with the
  performance reported in their papers.
We use Viper's publicly-available code~\cite{viper_code}.
We also evaluate \sysname against the following baselines:
  The peak PM bandwidth of a single \nvdimm of
    2.2 GB/s for sequential writes and
    5.85 GB/s for random reads.
With six \nvdimms in a NUMA node, the peak PM bandwidth of
    12.7 GB/s for sequential writes and
    31.5 GB/s for random reads.
% With 24 \nvdimms across 4 NUMA nodes, the peak PM bandwidth is
%     48.3 GB/s for sequential writes and
%     124.2 GB/s for random reads.

% \vheading{Peak throughput}.
% In our experiments, we compare against the peak
%   throughput of FlatStore, Viper, and ChameleonDB.
% We increase the number of application threads (up to 48 threads)
%   and report the maximum throughput of these PM stores.
% We observe that 8--16 application threads were sufficient to measure
%   their peak throughput on an interleaved PM device over
%   6 \nvdimms in a NUMA node.
% However, \sysname requires 4--8 application threads threads per \nvdimm
%   to avoid stalling for requests.
% Hence, we enqueue requests before hand to measure
%   the peak throughput of \sysname.
% Note that \sysname queues support
%   high, scalable enqueue throughput ($\approx$5 Mops/s per thread).

\vheading{Workloads and configurations}.
We use the YCSB benchmark suite~\cite{cooper2010benchmarking}
  which consists of workloads that represent real-world applications.
We generate YCSB workloads with 8B keys and 256B values (total size: 12.29 GB).
We use YCSB's LoadA (write-only) with 100\% writes,
  RunA (write-heavy) with 50\% reads and writes,
  RunB (read-heavy) with 95\% reads and 5\% writers,
  RunC (read-only) with 100\% reads,
  RunD (read-latest) and
  RunF (read-modify-write) workloads.
We generate YCSB workloads with uniform distribution and
  with moderate request skew using the default YCSB Zipfian
  coefficient of 0.99.
\sysname and PM stores we compare against do not support
  the RunE (range-scans) workload.
We compare \sysname against PM stores on a single \nvdimm and
  with 6 \nvdimms on a single NUMA node.
% We also compare \sysname against FlatStore for
%   workloads with small, varying value sizes using
%   2 nodes with 2 \nvdimms per node (FlatStore settings from their paper).
By default, \sysname uses 8 workers on a single \nvdimm. With six \nvdimms on a single node, \sysname uses 16 \workers per \nvdimm.
In our experiments, we use 8 application threads per \nvdimm by default.

% 16 workers, 8 clients	1nvdimm
% 8 workers per DIMM	6 nvdimms
% 32 clients	

\subsection{Throughput and Scalability}
\label{sec-eval:microbenchmarks}
We evaluate \sysname
  on a single \nvdimm and discuss its scalability
  to multiple \nvdimms across NUMA nodes.
We also evaluate \sysname on
  emulated CXL PM.
We use YCSB LoadA and RunC workloads with uniform distribution of keys for these experiments.

\vheading{Performance on a single \nvdimm}.
% Single \worker (trends and performance)
% Four \workers
% 8 \workers
% 16 \workers
\if 0
On a single \nvdimm and Load A workload, \sysname writes at 0.3 GB/s and obtains 14\% of PM write bandwidth using a single \worker.
\sysname processes about 1.2 Mops/s but requests have a lot of queueing delay (insufficient workers).
On scaling to 4 \workers per \nvdimm, we observe that
  \sysname obtains about 26\% (0.6 GB/s) PM write bandwidth;
  \sysname processes 2 Mops/s and takes 36us per request.
With 4 \workers, \sysname can handle up to 3 application threads without the latency of requests spiking up; \sysname processes 3 Mops/s, has 29us average latency, and obtains 40\% of PM write bandwidth.
With 4 \workers and 4 application threads, requests face queueing delays
  and \sysname obtains about 50\% of PM write bandwidth.
On increasing the number of \workers to 6,
  \sysname obtains 47\% of PM write bandwidth with 4 .
On increasing the number of application threads to 8 with 6 \workers, we see latency spikes.
With 8 \workers and 8 application threads , \sysname obtains up to 92\% PM write bandwidth
  but has 0.3s average request latency.
\fi
First, we configure the number of workers in \sysname for a single \nvdimm.
With 16 \workers and 8 application threads, \sysname obtains up to 88\% of PM write bandwidth.
\sysname processes 7 Mops/s and has average request latency of 53us.
With higher number of \workers, \nvdimms get overloaded
 which result in lower performance.
In this setting, \sysname can handle more than 48 application threads without drop in its throughput (Figure-\ref{fig:scalability-others}).
% Overall, \sysname achieves about 88\% of PM write bandwidth and processes about 7.2 Mops/s.
In the same setting, \sysname processes reads at about 6 Mops/s and has 30us latency,
  and obtains 25\% of PM read bandwidth.
\sysname uses 16 \workers, 1 background merger thread,
  and 16 application threads. This experiment uses 34 out of 54 available CPU cores 
  on a single NUMA node on the server.
Thus, the write throughput of \sysname is bandwidth-bound on a single \nvdimm and is CPU-bound within a NUMA node.
We now discuss the scalability of \sysname performance to \nvdimms across multiple NUMA nodes.

% 16 \workers/dimm	2x \workers	4x \workers
% 8 application threads / node	2x application threads	4x application threads
% NUMA	1 NVDIMM	2 NUMA nodes	4 NUMA nodes
% Uniform	Skye0	Skye2	skye4
% 	7	14	28
% 	8	8	5
% 	9	6	5
% 	9	3	3
% 	8	5	5
% 	9	8	6

\begin{table}[]
\begin{tabular}{@{}cccc@{}}
\toprule
\# NVDIMMs & Mops/s & Scaling & \% bw utilization \\ \midrule
1          & 7      & 1x      & 89\%              \\
2          & 14     & 1.9x    & 84\%              \\
4          & 28     & 3.9x    & 86\%              \\ \bottomrule
\end{tabular}
\mycaption{NUMA scalability}{
The write throughput of \sysname scales across multiple NUMA nodes. \sysname achieves 3.9\myx higher throughput on 4 nodes compared to one node from its fine-grained control over all PM accesses.}
\label{tab:numa-scaling}
\if\removespace 1
\vspace{-10pt}
\fi
\end{table}

\if 0
\begin{table}[]
\centering
\resizebox{\columnwidth}{!}{%
\begin{tabular}{@{}c|ccc|cc@{}}
\toprule
\multirow{2}{*}{\# NVDIMMs} & \multicolumn{3}{c|}{LoadA} & \multicolumn{2}{c}{RunC} \\ \cmidrule(l){5-6} 
 & Mops/s & Scaling & \% bw utilization & Mops/s & Scaling \\ \midrule
1 & 7 & 1x & 89\% & 9 & 1x \\
2 & 14 & 1.9x & 84\% & 6 & 70\% \\
4 & 28 & 3.9x & 86\% & 5 & 50\% \\ \bottomrule
\end{tabular}%
}
\end{table}
\fi
\begin{table}[]
\centering
\begin{tabular}{@{}l|cc@{}}
\toprule
Configuration               & CXL PM   & CXL PM (2\myx) \\ \midrule
\% write bw utilization     &   88\%       &   86\% \\
write (Mops/s)              &  7 (1\myx)    &   14  (1.9\myx) \\
write (GB/s)                &  1.9 (1\myx)    &  3.8 (1.9\myx)\\ \bottomrule
\end{tabular}
\mycaption{CXL PM}{
  \sysname throughput scales by 2\myx with two emulated CXL \nvdimms; \sysname utilizes up-to 88\% of the write bandwidth of emulated CXL PM.}
\label{tbl:cxl}
\if\removespace 1
\vspace{-20pt}
\fi
\end{table}

\vheading{Performance across multiple NUMA nodes}.
\sysname has scalable write throughput with increasing number of \nvdimms across NUMA nodes.
We scale \nvdimms by 4\myx across four NUMA nodes,
 and measure the write throughput of \sysname when it is
 configured to use 16 \workers per \nvdimm
  and 8 application threads per node.
Instead of a single \nvdimm in node-0,
when \sysname is configured to use two \nvdimms across two NUMA nodes,
\sysname processes 14 Mops/s and increases throughput by 94\%.
On configuring \sysname to use 4 \nvdimms across 4 NUMA nodes,
  \sysname processes around 28 Mops/s and achieves 3.9\myx higher throughput,
  as shown in Table-\ref{tab:numa-scaling}.
On a single \nvdimm,
 \sysname processes reads at 9 Mops/s;
 this read throughput does not scale
 with increasing number of \nvdimms.
On scaling to multiple NUMA nodes,
  \workers on \nvdimms remain underutilized due to
  load imbalance of read requests.
Note that with more \workers,
  \sysname increases its read bandwidth
  utilization.
% Thus read throughput in \sysname is latency-bound;
%   we observe the same trend in state-of-the-art PM stores (\sref{sec-eval:macro}).
% Hence \sysname performance drops to 6 Mops/s with 2 and 4 NUMA nodes with one \nvdimm per node.

\vheading{CXL Discussion}.
\label{sec-eval:cxl}
To evaluate the impact of PM capacity expansion on \sysname,
  we emulate CXL PM following CXL memory emulation
  from prior work~\cite{maruf2022tpp, ahn2022enabling}.
Note that recent CXL research~\cite{sun2023demystifying} reports that sequential
  accesses (using load and non-temporal store instructions) to
  remote NUMA memory of a dual-socket server have higher performance and
  lower latency than real CXL memory;
  we assume that this observation translates to PM.
% 
% \viheading{Baseline}.
We measure the peak write bandwidth of emulated CXL PM
  using the Intel Memory Latency Checker~\cite{intel_mlc}
  and our microbenchmarks from PM analysis (\sref{sec:Analysis}).
We observe that a single CXL-attached \nvdimm (CXL-PM)
  has a maximum write bandwidth of 2.23 GB/s (4 writers and 512B writes);
  CXL PM has similar sequential write bandwidth as PM.

% We evaluate how \sysname performs in this setup with NUMA-emulated CXL PM.
% We observe that
\sysname saturates 88\% of the available write bandwidth
  with CXL-PM on YCSB LoadA workload.
With CXL PM, \sysname processes $\approx$7 Mops/s and writes at 1.9 GB/s.
Further, with 2\myx the capacity of CXL PM, \sysname saturates
  86\% of available write bandwidth.
In this setting, \sysname processes $\approx$14 Mops/s and writes at
    3.8 GB/s.
Therefore, by increasing CXL-attached PM capacity by 2\myx,
  \sysname performance scales by 1.9\myx (Table~\ref{tbl:cxl}).
\sysname supports CXL-PM without any modifications and shows how
  the benefits of \sysname translate to persistent memory media
  with lower write bandwidth than PM.
Since CXL performance is not similar to remote NUMA performance for
  other access patterns like reads~\cite{sun2023demystifying},
  we only evaluate the performance of sequential writes
  (using non-temporal stores).

\vheading{Ablation study}.
We discuss end-to-end impact of the design choices of \sysname.
  % on the end-to-end throughput.
  % that implement the indirect-access architecture.
\sysname has up to 14\% lower throughput across all YCSB workloads
  when run on a single interleaved (six \nvdimms on a node) PM device instead
  of managing individual \nvdimms.
% On relying on PM memory controller to combine six \nvdimms
%   within a NUMA node into a single interleaved device
%   instead of managing one PM device per \nvdimm,
%   \sysname has 14\% lower performance on write-dominant workloads.
By managing non-interleaved \nvdimms, \sysname has higher benefits from hardware prefetching.
Next, \sysname uses fixed number of \workers per \nvdimms,
  implements indirect-access, and avoids overloading PM.
However, without indirect-access, too many applications threads can simultaneously access an \nvdimm and overload PM.
\sysname experiences 17\% lower performance on
  increasing the number of workers form 16 to 32 threads.
Further, using threads of a remote NUMA node
drops \sysname performance by up to 4\myx.
We note that \nvlog and  batching also provide significant performance improvements.  
% Overall, the write throughput is PM-bandwidth bound on a single \nvdimm and CPU bound with multiple \nvdimms within a NUMA node.
% the read throughput of \sysname is latency-bound and

% \begin{table}[]
% \centering
% \begin{tabular}{@{}cccccc@{}}
% \toprule
% Avg. (us) & FlatStore & Viper & ChameleonDB & \sysname & \sysname \\ \midrule
% Put & 3 & 7 (1\myx) & 4 & 35\  (5\myx) & 35\  (5\myx) \\
% Get & 3 & 3 (1\myx) & 3 & 32 (10\myx) & 32 (10\myx)\\ \bottomrule
% \end{tabular}
% \mycaption{Average request latency}{This table reports the
% average put and get request latency of PM key-value stores.}
% \label{tbl:put-get-latency}
% \end{table}

% \begin{table}[]
% \centering
% \resizebox{\columnwidth}{!}{%
% \begin{tabular}{lccccc}
% \hline
% \begin{tabular}[c]{@{}l@{}}Mean\\ Latency\end{tabular} & FlatStore & Viper & ChameleonDB & \begin{tabular}[c]{@{}c@{}}\sysname\\ (w/o batching)\end{tabular} & \sysname \\ \hline
% Put (us) & 3 & 7 (1\myx) & 4 & 3 & 35 (5\myx) \\
% Get (us) & 3 & 3 (1\myx) & 3 & 3 & 32 (10\myx) \\ \hline
% \end{tabular}
% }
% \mycaption{Average request latency}{This table reports the
% average put and get request latency of PM key-value stores.}
% \label{tbl:put-get-latency}
% \if\removespace 1
% \vspace{-20pt}
% \fi
% \end{table}

% \usepackage{multirow}
\begin{table}[]
\centering
\resizebox{\columnwidth}{!}{%
\begin{tabular}{|l|l|l|l|cc|}
\hline
\multirow{2}{*}{\begin{tabular}[c]{@{}l@{}}Mean \\ Latency\end{tabular}} &
  \multirow{2}{*}{FlatStore} &
  \multirow{2}{*}{Viper} &
  \multirow{2}{*}{ChameleonDB} &
  \multicolumn{2}{c|}{\sysname} \\ \cline{5-6} 
         &                        &                             &                        & \multicolumn{1}{l|}{No Batching} & \multicolumn{1}{l|}{Batching} \\ \hline
Put (us) & \multicolumn{1}{c|}{3} & \multicolumn{1}{c|}{7 (1\myx)} & \multicolumn{1}{c|}{4} & \multicolumn{1}{c|}{3}           & 35 (5\myx)                       \\ \hline
Get (us) & \multicolumn{1}{c|}{3} & \multicolumn{1}{c|}{3 (1\myx)} & \multicolumn{1}{c|}{3} & \multicolumn{1}{c|}{3}           & 32 (10\myx)                      \\ \hline
\end{tabular}
}
\mycaption{Average request latency}{This table reports the
average put and get request latency of PM key-value stores.}
\label{tbl:put-get-latency}
\if\removespace 1
\vspace{-20pt}
\fi

\end{table}

\subsection{Latency}
\label{sec-eval:put-get-latency}

We now discuss the latency characteristics of \sysname.
We use default YCSB LoadA and RunC workloads for these experiments.

\vheading{Low latency settings}.
Workers process requests in batches by default;
workers wait till a configurable number of requests arrive at their request pending queues before processing them.
While batching increases
  the average latency of requests, it enables high throughput;
  \sysname achieves a peak write throughput of 7.2 Mops/s (88\% of PM write bandwidth) on a single \nvdimm (\sref{sec-eval:microbenchmarks}).
Further, batching reduces CPU consumption and limits the
  contention on the request pending queues.
If applications disable batching then
  \sysname takes around 3us to process a put or a get
  request and has comparable latencies with the state-of-the-art PM stores,
  as shown in Table-\ref{tbl:put-get-latency}.
Without batching, \sysname has 30\% lower peak write throughput; it processes 5.4 Mops/s on a single \nvdimm (66\% of PM write bandwidth).

\begin{figure}
\centering
\small
\includegraphics[width=\columnwidth]{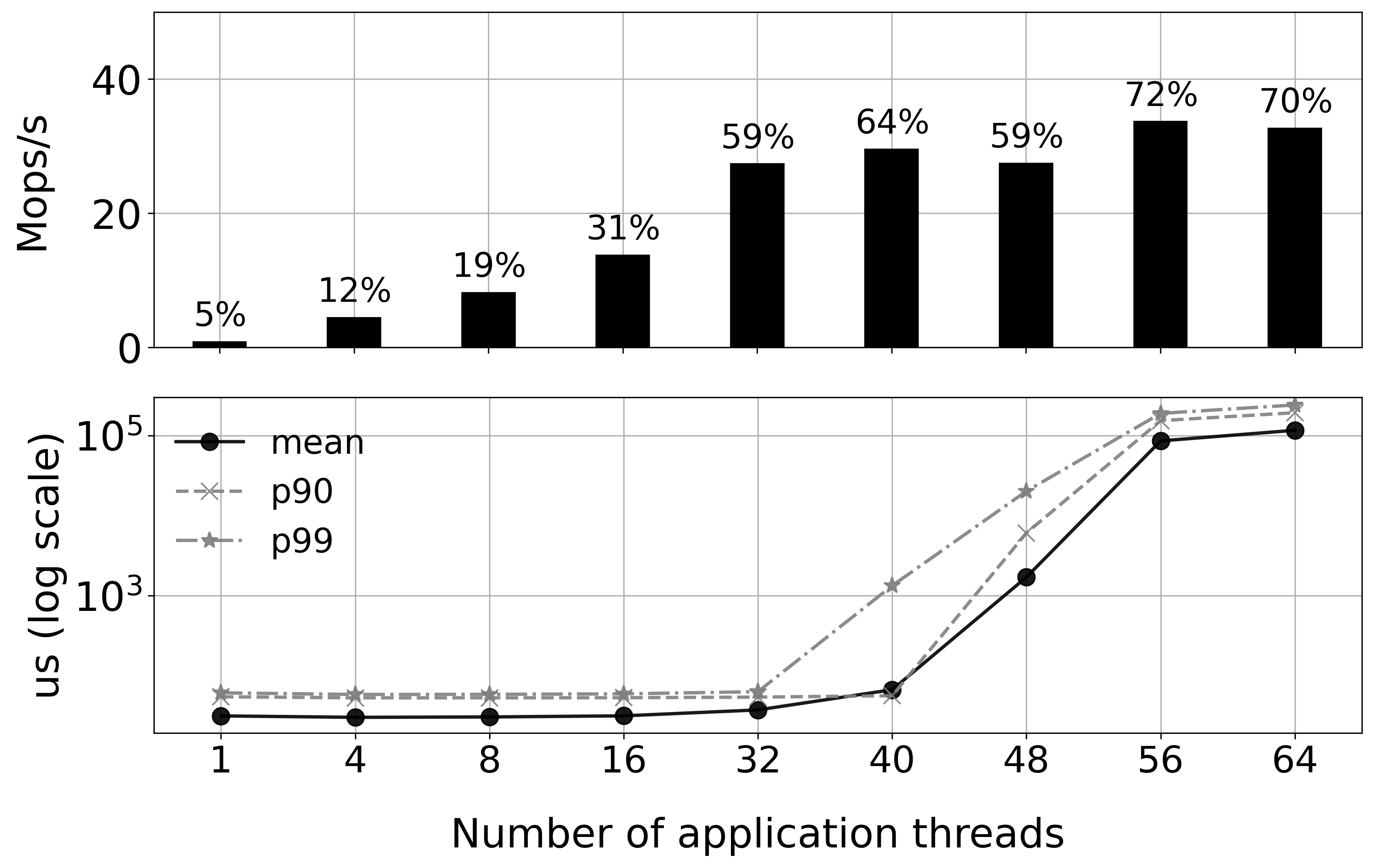}
\if\removespace 1
\vspace{-20pt}
\fi
\mycaption{Throughput and latency}{The throughput of \sysname increases
  with the number of application threads, it saturates with
  $>$48 threads; latency continues to increase as request wait longer in the
  queues. Labels show PM bw utilization; \sysname uses 48 workers (8 per \nvdimm).}
% This figure shows \sysname's throughput and latency for put requests on a single NUMA node with 6 NVDIMMs and 8 workers per NVDIMM. The labels on the bars indicate PM write bandwidth utilization. \sysname is able to achieve 64\% bandwidth utilization with a latency less than 100 us for 40 clients. On further increasing the number of clients, the latency spikes significantly, while the throughput stagnates at 70\%.}
\label{fig:throughput-latency}
\if\removespace 1
\vspace{-10pt}
\fi
\end{figure}

\vheading{Put latency}. With 32 application threads (high load) and in default settings,
\sysname takes $\approx$35us to process a put request.
Application threads take 16\% of this time to receive a serial number and to get routed to a \worker's request queue.
They spend $\approx$44\% of the time to enqueue the requests, due to overheads from the concurrent queues.
% (Fig-\ref{fig:queue-latency}).
The \workers in \sysname take $\approx$39\% of the time to dequeue to the request,
  append data and metadata to the logs on \nvdimms,
  and update the in-memory merge log.
% We observe that PM write latency is similar to DRAM write latency but
%   request enqueuing latency is 5\myx higher than the dequeue latency
%   from the lock-free request pending queues.

\vheading{Get latency}. Similarly, with high load,
\sysname takes $\approx$32us to process a get request.
\sysname spends $>$90\% of this time to enqueue the request
  and to notify the \worker which dequeues the request.
  The \worker takes around 4us ($<$10\% of the time)
  to read the value from data logs on PM and to set the
  return value and its expected checksum.
In the fast path, application threads cache the
  metadata from put requests and forward it along with
  the get requests.
However, if application threads do not provide metadata
  (\eg serial number of the key),
  then \sysname searches the merge logs and indexes
  for the latest value of the key and
  routes the application threads to \workers;
\sysname takes $\approx$60us to process the request
  in its slow path.

\vheading{Tradeoffs}.
There is a fundamental tradeoff between providing low latency to applications and obtaining
high throughput.
Indirect-access is crucial for achieving high throughput (limits the number of concurrent accesses to \nvdimms),
however, it requires additional work (like dequeuing requests) in
  the critical path and increases the latency of operations.
% Further, performing additional work requires having more threads
%   to operate at close-to PM write bandwidth; thus it
%   requires more CPU threads.
% Thus, \sysname provides indirect-access for applications, trades off low latency and obtains high write throughput; \sysname obtains high PM write bandwidth utilization and provides stable throughput to applications with increasing number of threads.
%Thus, \sysname provides indirect-access for applications,
 % obtains high throughput %and trades off low latency;
  %\sysname obtains high PM write bandwidth utilization
  %and provides stable throughput 
%  with increasing number of application threads and trades off low latency.

\vheading{Throughput and latency discussion}.
% \label{sec-eval:micro}
\sysname is designed for applications that require high throughput and can tolerate microsecond-scale latencies.
For applications that can tolerate even higher latencies,
  \sysname provides a throughput and latency tradeoff,
  as shown in Figure-\ref{fig:throughput-latency}.
Applications can scale their threads based on their latency SLAs.
On a single NUMA node with six \nvdimms and with 32 application threads, \sysname processes $\approx$28 Mops/s 
  with 8 \workers per \nvdimm.
\sysname saturates $\approx$59\% of PM write bandwidth while taking
  $\approx$37us to process a put request on average.
In this configuration,
  \sysname has a 99\% tail latency of 63us.
For applications that benefit from higher throughput and can tolerate higher latencies, \sysname allows applications to scale their threads and provides higher performance.
With 56 application threads, \sysname processes 34 Mops/s with an average latency of 85ms per request and a 99\% latency of 0.2s respectively,
% With 56 application threads and 8 \workers per \nvdimm, \sysname 
and obtains $\approx$72\% of PM write bandwidth.

\begin{figure}[t]
\centering
\includegraphics[width=\columnwidth]{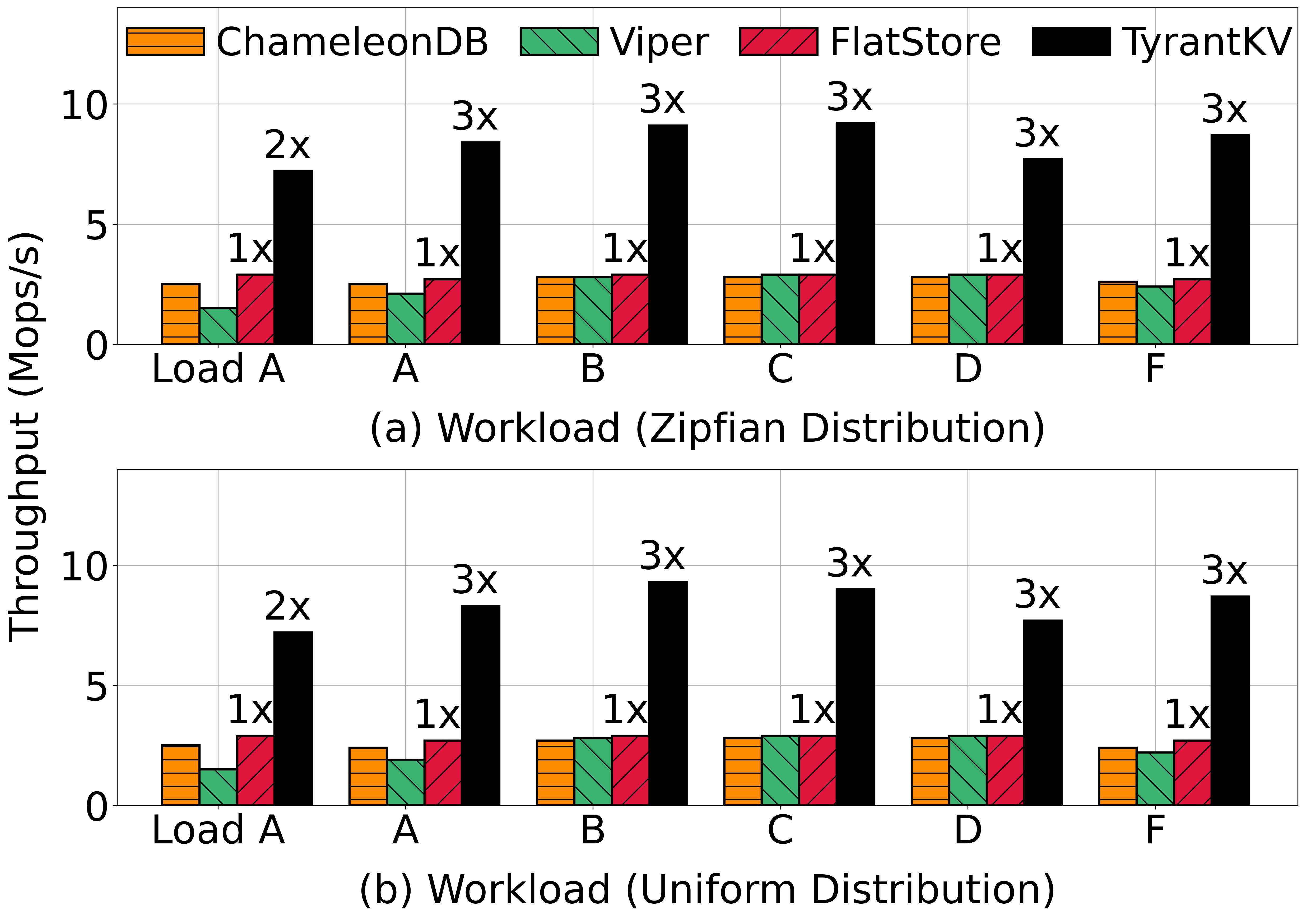}
\if\removespace 1
\vspace{-20pt}
\fi
\mycaption{Comparison against PM stores on a single \nvdimm}{
  \sysname outperforms FlatStore
  for all YCSB workloads by up to 3\myx; this is from providing indirect-access for application threads and using logs from \nvlog.}
\label{eval:ycsb-single-dimm}
\if\removespace 1
\vspace{-10pt}
\fi
\end{figure}

\subsection{Yahoo Cloud Serving Benchmark}
\label{sec-eval:macro}
% We evaluate the performance of \sysname using YCSB that contains workloads representing
% real-world application access patterns.
We compare \sysname to prior PM stores.
For each experiment, we load the PM stores with 50M key-value pairs and run YCSB workloads with 50M operations.
The state-of-the-art PM stores run on a single interleaved PM device across six \nvdimms in a NUMA node. Since \sysname is configured to use 16 \worker threads, other PM stores use 16 application threads; \sysname uses 8 application threads to provide 30--35us latencies. We use the same number of threads to access PM across all PM stores; with higher number of application threads \sysname has higher throughput.

\begin{figure}[t]
\centering
\includegraphics[width=\columnwidth]{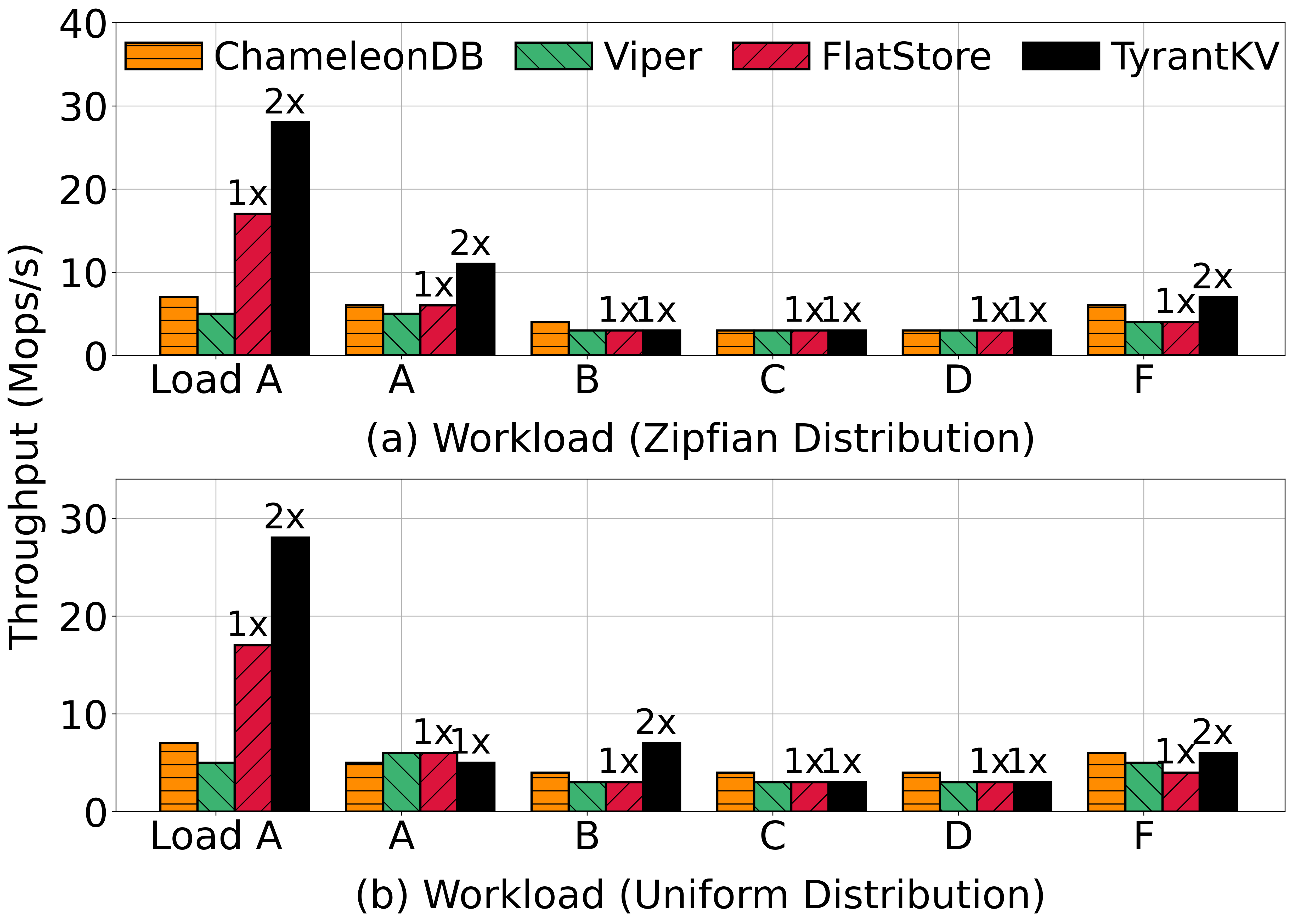}
\if\removespace 1
\vspace{-20pt}
\fi
\mycaption{Comparison against PM stores on a single node}{
  \sysname outperforms FlatStore by 2\myx on LoadA
  and performs comparably for read-dominant workloads with six \nvdimms on one node;
  the \workers in \sysname are underutilized due to low load.}
\label{eval:ycsb-single-node}
\if\removespace 1
\vspace{-15pt}
\fi
\end{figure}

\vheading{Single \nvdimm}.
% Figure~\ref{eval:ycsb-single-dimm} shows the performance of PM stores on a single \nvdimm for workloads with Zipfian (a) and Uniform (b) distributions. 
\sysname outperforms state-of-the-art PM key-value stores by 3--4\myx on all YCSB workloads and achieves a throughput of at least 8.4 Mops/s for workloads with Zipfian distribution (Figure~\ref{eval:ycsb-single-dimm}(a)) and about 7.2 Mops/s for workloads with Uniform distribution (Figure~\ref{eval:ycsb-single-dimm}(b)).
We observe that \sysname has higher performance for Zipfian distributions relative to Uniform for read-dominant workloads; the request skew helps workers utilize higher PM read bandwidth.

\viheading{LoadA}.
% vc: not useful to simply report numbrs
%\sysname processes put requests at 7.2 Mops/s.
%FlatStore, Viper, and ChameleonDB have a throughput of 2.9 Mops/s, 1.5 Mops/s, and 2.5 Mops/s respectively.
\sysname outperforms current PM stores by 2\myx on write-only workloads.
% \rohan{Unnecessary: \sysname has an average latency of 53us while the other key-value stores have 2--3us average latency.} 
\sysname trades off low latency for high and scalable PM write bandwidth utilization.
\sysname obtains about 88\% of PM write bandwidth in this configuration and scales to multiple application threads without overloading PM due to its novel indirect-access architecture.

\viheading{RunA}.
\sysname achieves 8.4 Mops/s on RunA workloads with Zipfian distribution and about 8.3 Mops/s for workloads with uniform distribution.
FlatStore, Viper, and ChameleonDB obtain 1.9--2.7Mops/s.
\sysname outperforms these PM stores by 3\myx on the write-dominant RunA workloads; \sysname achieves better bandwidth utilization for writes with \nvlog and its indirect-access architecture.

\viheading{RunB}.
\sysname processes 9.1 and 9.3 Mops/s for RunB workloads with Zipfian and Uniform distributions.
\sysname outperforms FlatStore by 3\myx that achieves 2.9 Mops/s.
\sysname outperforms ChameleonDB and Viper that processes 2.8 Mops/s by 3.3\myx. With 16 workers on a single \nvdimm, \sysname utilizes the available read bandwidth better than existing PM stores (using dedicated workers) and hence achieves better read throughput.

\viheading{RunC}.
\sysname processes the read-only RunC workload at 9 Mops/s. It outperforms FlatStore, Viper, and ChameleonDB by 3\myx as they process 2.8--2.9 Mops/s. On a single \nvdimm with 16 \workers and 8 application threads, \sysname performance for reads is latency bound, however, it uses available PM read bandwidth better than existing PM stores (by using dedicated workers).
\sysname obtains about 25\% of PM read bandwidth in this workload. With higher number of \workers, we observe higher read throughput for \sysname and better PM read-bandwidth utilization. However increasing the number of \workers trades off write performance and increases CPU utilization.

\viheading{RunD}.
\sysname processes workloads with read-latest distribution at 7.7 Mops/s. Other PM stores obtain around 2.8--2.9 Mops/s. \sysname outperforms them by 3\myx. With skewed reads, \sysname achieves
better read bandwidth utilization and hence outperforms PM stores.

\viheading{RunF}.
\sysname obtains a throughput of 8.7 Mops/s
  with RunF workload.
FlatStore, Viper, and ChameleonDB obtain 2.2--2.7 Mops/s. \sysname outperforms these PM stores by 3\myx. \sysname does not support special read-modify-write operations. Hence, the performance of \sysname for RunF workload is similar to workloads with similar put and get request distributions.

\viheading{Summary}.
% Overall, we observe that
The state-of-the-art PM stores allow application threads to directly access PM and suffer from poor throughput and scalability. \sysname uses dedicated threads to access the \nvdimm and uses efficient \nvlog to obtain high read and write throughput. \sysname supports higher number of application threads without a drop in throughput. We show that
designing for high PM bandwidth utilization
provides better read and write throughput. 
% \sysname trades off low latency and achieves high performance for both puts and gets due to its PM-conscious design.

\vheading{Single NUMA node}
We evaluate \sysname on a single
NUMA node with six \nvdimms.
We observe that \sysname has at least 2\myx higher performance compared to FlatStore on LoadA and has comparable performance on run workloads, with Zipfian (Figure~\ref{eval:ycsb-single-node}(a)) and Uniform (Figure~\ref{eval:ycsb-single-node}(b)) distributions.

\viheading{Write-dominant workloads}.
% We observe that 
\sysname outperforms state-of-the-art PM stores like FlatStore by 2\myx, Viper by 4\myx, and ChameleonDB by 5\myx, on the write-only LoadA workload.
For write-heavy RunA workloads, \sysname outperforms other PM stores by 2\myx.

\viheading{Mixed workloads}.
With mixed workloads, PM stores and \sysname have much lower throughput relative to their throughput on a single \nvdimm. 
The PM read bandwidth is twice that of the write bandwidth while the latency is 2-3\myx that of the write latency. Therefore, accessing PM with 16 threads does not saturate its bandwidth; hence read throughput for all PM stores becomes latency-bound. In this setting, \sysname has comparable performance to other PM stores.
However, with higher number of \workers, \sysname achieves better read throughput since read bandwidth is better utilized with more threads. Thus, with multi-\nvdimm per NUMA node settings \sysname read throughput is CPU-bound.

% \skl{I could not clearly understand the reason why PM stores have lower read throughput with more \nvdimms than a single \nvdimm setup. Does it mean that we didn't run the workloads with enough loads for utilizing additional PM read bandwidths with more \nvdimms?}

% under low load, read performance of all PM stores is latency bound. \sysname achieves comparable performance on read workloads, as shown in Figure~\ref{eval:fs-viper-comp}.

% Since PM write bandwidth is not saturated, \sysname observes higher write throughput.

\subsection{Tradeoffs and Overheads}
\label{eval:overheads}

\vheading{Memory utilization}.
\sysname uses $\approx$8 GB of DRAM for 768 GB of PM per NUMA node (1\%). 
\sysname uses FASTER for its in-memory index per NUMA node and maintains
  1GB-sized data and metadata logs on PM.
We observe that, a majority of \sysname's total memory footprint
  is from request queues; \sysname uses per-\worker read and write pending queues and merge logs.
\sysname uses per-\worker structures and utilizes more memory to avoid concurrency overheads in the critical path; scaling them further based on the number of application threads reduces these overheads at the cost of more memory.
% Thus, \sysname keeps memory utilization in check while achieving high throughput and PM write bandwidth utilization.

\vheading{CPU utilization}.
% \sysname makes interesting tradeoffs that impact the CPU utilization, throughput, and latency of requests.
\sysname builds on the insight that concurrent \vio to \nvdimms
  provides similar throughput with fewer threads relative
  to avoiding them (contrast to prior recommendations). 
% \sysname requires 16 workers to achieve 88\% of PM write bandwidth utilization on a single \nvdimm.
If \sysname were to achieve high PM random-read bandwidth, it would require 4\myx higher CPU consumption;
\sysname trades off read performance for lower CPU utilization and higher write throughput.
\sysname performs concurrent reads and writes on \nvdimms to achieve higher throughput with fewer number of workers per \nvdimm.
Further, workers in \sysname sleep till a fixed (configurable) number of requests arrive at their queues; this helps lower utilize CPU utilization and increases write throughput.

\vheading{Recovery time}.
% We observe that 
During recovery, \sysname first reads records from data and metadata logs.
Then its parses the records which takes 2.7\myx more time than sequentially reading from PM logs,
  and then reconstructs the in-memory structures which takes 15\myx more time.
Overall, \sysname recovers at 1.7 GB/s per NUMA node and is limited by 
the performance of in-memory metadata indexes.
\subsection{Performance Impact of Tunable Parameters}
\label{sec-eval:tunable}

% \begin{table}[]
% % \resizebox{\columnwidth}{!}{%
% \begin{tabular}{@{}llllll@{}}
% \toprule
% \#Clients & 1   & 2   & 4   & 6   & 8     \\ \midrule
% Mops/s        & 2.3 & 3.4 & 4.6 & 5.7 & 6.4   \\
% Avg(us)       & 33  & 34  & 31  & 43  & 75383 \\ \bottomrule
% \end{tabular}%
% % }
% \mycaption{\sysname with single \worker}{This table shows \sysname performance
% on a single NUMA node with six \nvdimms for the LoadA workload with increasing
% number of clients.}
% \label{eval-tbl:single-worker}
% \end{table}

% \begin{table}[]
% % \resizebox{\columnwidth}{!}{%
% \begin{tabular}{@{}llllll@{}}
% \toprule
% \#Clients & 1   & 4   & 8   & 16   & 32    \\ \midrule
% Mops/s        & 1.6 & 3.6 & 4 & 3.3 & 2   \\
% Avg(us)       & 29  & 29  & 107  & 36  & 47 \\ \bottomrule
% \end{tabular}%
% % }
% \mycaption{\sysname with 8 \workers}{This table shows \sysname performance
% on a single NUMA node with six \nvdimms for the LoadA workload with increasing
% number of clients.}
% \label{eval-tbl:single-worker}
% \end{table}

% \usepackage{booktabs}
\begin{table}[]
\begin{tabular}{@{}l|ccccc|cccc@{}}
\toprule
Workload  & \multicolumn{5}{c|}{Puts} & \multicolumn{4}{c}{Gets}  \\ \midrule
$\#$App. threads & 1   & 2   & 4   & 6   & 8   & 1   & 4  & 16  & 32 \\ 
Mops/s    & 2 & 3 & 5 & 6 & 6 & 2 & 4  & 3 & 2  \\
% Mops/s    & 2.3 & 3.4 & 4.6 & 5.7 & 6.4 & 1.6 & 3.6 & 4   & 3.3 & 2  \\
Avg Latency (us)    & 33  & 34  & 31  & 43  & 75k & 29  & 29  & 36  & 47 \\ \bottomrule
\end{tabular}
\mycaption{One \worker per \nvdimm}{The throughput and mean latency of put and get ops in \sysname with one worker
per \nvdimm. The worker waits to batch under low load, and requests wait for longer in the queues with $>$6 app. threads.}
\label{tbl-single-worker}
\if\removespace 1
\vspace{-15pt}
\fi
\end{table}

%\sysname provides indirect-access for applications using dedicated \emph{\worker} threads.
% \sysname allows applications
%   to configure the number of \worker threads per \nvdimm; this design
%   enables high throughput with increasing number of application threads.
% We now evaluate the performance impact of configuring
%  the number of \workers per \nvdimm with increasing number
% of application threads.
% We use the default YCSB LoadA and RunC  workloads and run \sysname
% on a single NUMA node with six \nvdimms.

\sysname uses dedicated workers to provide indirect-access
  and allows applications to configure the number of workers.
We discuss tuning the number of workers and its impact
  on the throughput and latency of \sysname
  while increasing number of application threads.
We use the default YCSB LoadA and RunC  workloads
  and run \sysname on a single NUMA node with six \nvdimms.

% \begin{figure}
% \centering
% \small
% \includegraphics[width=\columnwidth]{evaluation/new-graphs/fig-latency-bandwidth-multiworker.png}
% \if\removespace 1
% \vspace{-20pt}
% \fi
% \mycaption{\sysname with 8 workers}{This table shows \sysname performance
% on a single NUMA node with six \nvdimms for the LoadA workload with increasing
% number of clients.}
% \label{fig:8-worker-write}
% \if\removespace 1
% \vspace{-10pt}
% \fi
% \end{figure}

\begin{figure}
\centering
\small
\includegraphics[width=\columnwidth]{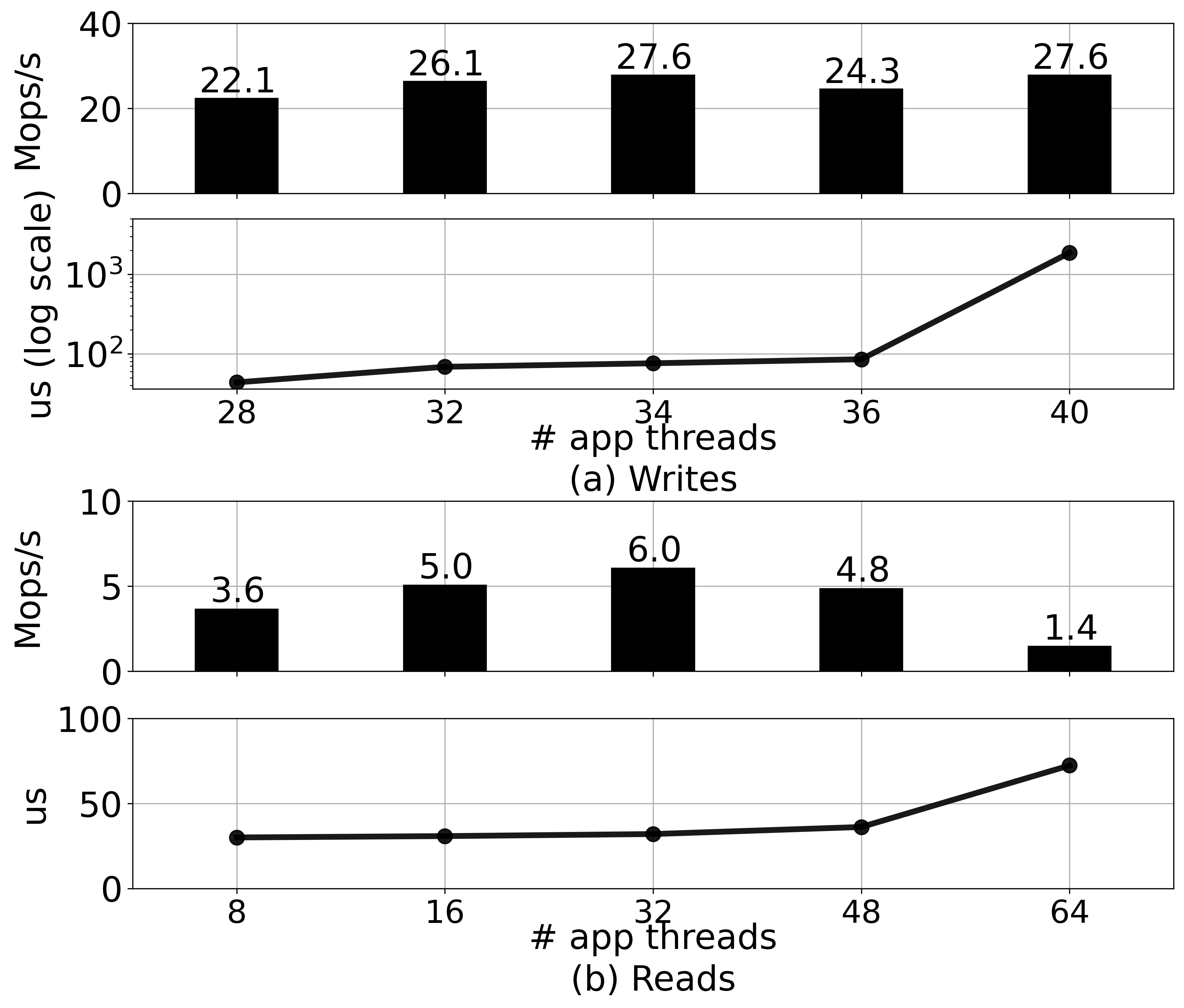}
\if\removespace 1
\vspace{-20pt}
\fi
\mycaption{\sysname with 8 workers per \nvdimm}{The throughput and mean latency for put (a) and get (b) requests in \sysname with increasing $\#$app. threads. The write throughput of \sysname is bound by PM bandwidth which causes writes to wait longer in queues with $>$40 app. threads (note the log scale); the read throughput of \sysname is CPU-bound and latency increases gradually.}
\label{fig:8-worker-rw}
\if\removespace 1
\vspace{-10pt}
\fi
\end{figure}

\vheading{Write performance}.
With a single application thread
  and when configured to use one \worker per \nvdimm,
  \sysname has a write throughput of $\approx$2 Mops/s.
On scaling applications to six threads,
  write throughput of \sysname gradually increases to $\approx$6 Mops/s.
With 1--6 application threads,
  \sysname observes an average put latency of 31--43us,
  as shown in the Table-\ref{tbl-single-worker}.
However, with 8 application threads, while write throughput of \sysname increases to 6.4 Mops/s,
  the average latency of requests shoots up to 75ms.
We observe that the sudden spike in latency is due to queueing delays;
  requests wait in the worker's queues for longer.
Thus, \sysname needs more \workers to handle beyond six application threads.

\vheading{Scaling \workers}.
Increasing the number of \workers improves
  the write throughput and latency of \sysname.
With two \workers per \nvdimm and  8 application threads,
  \sysname processes $\approx$9 Mops/s and takes 32us per request.
% On scaling applications to 10 threads,
%   we observe that the average latency of requests
%   has a sharp spike due to queueing delays.
With two workers per \nvdimm, \sysname
  can handle $<$10 application threads before requests experience high queueing delays.
On increasing the number of \workers to four per \nvdimm,
  \sysname handles up to 24 application threads.
% Thus, we increase the number of \workers to four per \nvdimm and observe
%   that \sysname can handle up to 24 application threads;
%   \sysname processes $\approx$20 Mops/s with $\approx$76 us average put latency.
% To handle more ($>$24) application threads,
%   we configure \sysname with 8 \workers per \nvdimm.
With 8 \workers per \nvdimm,
  \sysname process 27.6 Mops/s and scales till 34 application threads,
  as shown in Figure-\ref{fig:8-worker-rw}(a).
We observe that increasing the number of application threads further does not improve the write throughput of \sysname.
Even with 40 application threads, write throughput remains stable at 27.6 Mops/s
  while the average request latency spikes as requests
  wait in the queues for longer (Figure-\ref{fig:8-worker-rw}(b); please note the log scale).
  % due to queueing delays (please note the log scale in the graph).
%Thus, \sysname provides high throughput with increasing number
 % of application threads and couteracts PM sensitivities (too many threads lower performance) with its indirect-access architecture.

% \input{figures/new-figures/fig-queue-latency}

\vheading{Read performance}.
With a single application thread and
when configured to use one \worker per \nvdimm,
  \sysname processes reads at $\approx$2 Mops/s and takes
  $\approx$30us to process a get request on average (Table-\ref{tbl-single-worker}).
Scaling application (till 16 threads) improves the read throughput of \sysname.
% With four application threads,
%   the read throughput of \sysname increases to $\approx$4 Mops/s.
% However, on scaling applications to 16 threads,
%   \sysname has lower read throughput of $\approx$3 Mops/s and takes
%   37us on average to process get requests.
% Thus, we increase the number of workers per \nvdimm
%   and measure the read performance of \sysname.
With higher number of workers (8 per \nvdimm)
  and with 32 application threads,
  \sysname achieves a read throughput of 6 Mops/s and takes
  32us per request.
Beyond 32 application threads, \sysname's read throughput decreases; note that the average request latency grows slowly,
  as shown in Figure-\ref{fig:8-worker-rw}(b).
This behavior is primarily due to the poor scalability of queues.
  % enqueuing throughput
  % of application threads due to \vio bottlenecks from
  % the concurrent request pending queues.
  % in the communication fabric (Fig~\ref{fig:queue-latency})
We observe that with 32 concurrent threads, the lock-free
  queue~\cite{libboost} has an average enqueue and dequeue latency of 23us.
Even with a single thread, queues add an extra 2us latency
  while enqueueing and dequeueing requests (note: PM latency is $\approx$100--300ns).
Thus, we need low-latency high-throughput concurrent queues
  to fully realize the benefits of indirect-access architectures.

\vheading{Write-optimization and tradeoffs}.
The read throughput of \sysname grows gradually
  relative to its write throughput with increasing number of workers; we observe similar trends in all PM stores (\sref{sec-eval:macro}).
This is fundamentally due to the higher read latency of PM;
  PM read latency is almost 2--3\myx higher than its write latency.
Our analysis shows that PM requires up to 4--6\myx higher number of threads to
  effectively utilize PM read bandwidth relative to its write bandwidth (\sref{sec:Background}).
However, increasing the number of \workers to increase read throughput
  trades off write throughput (\sref{sec:Analysis}).
Hence, \workers in \sysname are configured to achieve
  high write throughput by default; this trades off read throughput and reduces CPU utilization.

\section{Related Work}
\label{sec:Related-Work}

%We place our contributions in the context of related prior work.

\vheading{PM key-value stores}.
Key-value stores designed from scratch for PM~\cite{benson2021viper, chen2020flatstore,wang2022pacman} exploit the byte-addressability and the low latency of PM. 
The work closest to \sysname is Viper~\cite{benson2021viper}.
Viper evenly distributes threads across NVDIMMs and designs NVDIMM-aligned storage layouts to benefit from the parallelism of interleaved PM and avoid cross-NVDIMM contention.
However, prior work does~\cite{chen2020flatstore,zhang2021chameleondb,benson2021viper} not take into account the limited concurrent access allowed by PM, or that managing individual \nvdimms yields higher throughput.
Finally, prior work~\cite{zhang2021chameleondb,benson2021viper} is not designed to be NUMA-aware, exhibiting poor performance (or lack of support) on multiple NUMA nodes. 
In contrast, \sysname takes all these aspects into consideration and implements the indirect-access for applications, and obtains high and scalable write throughput.
% ; note, \sysname performs comparably to other PM stores on read-dominant workloads.

\vheading{OdinFS}.
A concurrent work that is close to \sysname in the file-system domain
  is OdinFS~\cite{zhou2022odinfs}.
\sysname and OdinFS share several goals: scalability, NUMA-awareness,
  and PM bandwidth utilization.
OdinFS improves performance for syscall-based applications.
Since it relies on stripping to scale across nodes,
  it trades off PM contiguity and hurts the performance of \mmap applications.
\sysname supports a simpler key-value API,
  uses \mmap interface, has a modular design,
  and obtains high PM write-bandwidth utilization.
\sysname exploits the benefits of indirect-access (in contrast to current PM stores) without paying the overheads from POSIX API unlike OdinFS.

\if 0
\vheading{OdinFS}. 
A concurrent work that is close to \sysname in the file-system domain is OdinFS~\cite{zhou2022odinfs}.
\sysname and OdinFS share several goals include scalability, NUMA-awareness, and bandwidth utilization. 
Similar to \sysname, OdinFS also uses delegation to perform I/O, rather than allowing application threads to perform I/O directly. 
OdinFS uses "controlled
PM access approach", similar to how \sysname limits concurrent access. 
However, to the best of our knowledge, 
OdinFS does not perform I/O in read and write phases; it is possible this is harder to realize in a file system than in a simpler key-value store. 
Apart from this difference, though both \sysname and OdinFS have the same goals, the details of the designs turned out to be quite different owing to the difference between the more complex file systems and simpler key-value stores. 

\vheading{Software-Defined PM}.
Software-defined PM shares similarities with prior work on Software-defined networking (SDN)~\cite{SDN} and Software-defined storage (SDS)~\cite{SDS}; all these efforts aim to abstract an underlying hardware with standardized APIs and virtualize it, thereby allowing user applications to deploy the hardware with greater flexibility, functionality, and scalability. Similarly, software-defined PM abstracts PM media complexities by simplifying the usage with the standardized interface and allows scale-out by employing PM optimizations under the hood. 
\fi

\section{Conclusion}
\label{sec:Conclusion}

We present \sysname, a novel write-optimized PM key-value store.
The core idea behind \sysname is to retain fine-grained control over all PM accesses;
\sysname provide indirect-access to applications and manages individual \nvdimms
to obtain high and scalable PM write-bandwidth utilization.
We demonstrate that \sysname outperforms state-of-the-art PM stores
  in different settings and across various workloads.
In the future, we envision \sysname supporting newer PM media.
We plan to provide open-access to our prototype implementation of \sysname. 
% Add we discuss the practical implications of designing a high-throughput PM key-value store and present its fundamental tradeoff with CPU utilization and low latency!

\if 0
We present \sysname, a novel key-value store for PM that 
achieves scalability and high performance. 
The core idea behind \sysname is identifying the I/O bottlenecks of PM, 
and working around those bottlenecks with software techniques. 
By carefully controlling how I/O is performed, 
we demonstrate that \sysname is able to significantly increase
 utilization of PM write bandwidth. 
We plan to make the \sysname prototype publicly available. 

The emerging CXL is a new technology that helps expand
  the available PM capacity per server and makes PM a
  promising technology for building large-scale,
  persistent key-value stores.
Prior PM stores sacrifice performance for simple PM media management.
We reiterate four design principles that are crucial for achieving
  high PM write bandwidth utilization, a limited PM resource.
We present \sysname, a new PM store, that isolates and enables
  PM-optimal design with \spmfull, and follows all four principles
  with its data and control plane architecture and a modular design.
\sysname achieve high PM write bandwidth utilization
  that scales with available PM capacity,
  which we demonstrate through empirical evaluation.

\fi

% \section*{ACKNOWLEDGEMENTS}
% This work was supported in part by Semiconductor Research Corporation (SRC).

% \input{intro}
% \input{background}
% \input{analysis}
% \input{design}
% \input{impl}
% \input{eval_osdi}
% \input{eval}

% \clearpage

\bibliographystyle{ACM-Reference-Format}
\bibliography{main}

\end{document}